\documentclass[11pt]{iopart}
\usepackage{graphicx}
\usepackage{epstopdf}
\DeclareGraphicsExtensions{.pdf,.jpg,.png,.eps,.ps}
\usepackage{amsfonts}
\expandafter\let\csname equation*\endcsname\relax
\expandafter\let\csname endequation*\endcsname\relax
\usepackage{amsmath}
\usepackage{color}

\begin{document}
\nocite{*}
\title[Multi-sublattice magnetism and superexchange interactions in double-double perovskite CaMnCrSbO$_{6}$]{Origins of multi-sublattice magnetism and superexchange interactions in double-double perovskite CaMnCrSbO$_{6}$}
%
\author {Rakshanda Dhawan$^{1,\ddagger}$, Padmanabhan Balasubramanian$^{2,\dagger,\ddagger}$, Tashi Nautiyal$^1$}
\vspace{5 mm}
{\small $^1$Department of Physics, Indian Institute of technology, Roorkee-247667, Uttarakhand, India.
\newline
$^2$Department of Physics, Graphic Era University, Dehra Dun, Uttarakhand 248 002, India.}
\ead{$^{\dagger}$bpaddy123@gmail.com}
\footnote{Both the authors have made equal contribution.}
\begin{abstract}
The recently reported double-double perovskite compound CaMnCrSbO$_{6}$ offers an interesting possiblity for the theoretical exploration of exchange interactions between the multiple magnetic sublattices in the system. We have deployed density functional theory, Wannier function analysis and mean-field calculations to investigate this compound. The crystallographically non-equivalent Mn atoms in the unit cell have tetrahedral and planar oxygen coordinations (labelled as Mn(1) and Mn(2)), while the Cr atom is in the centre of distorted oxygen octahedron.
While the bulk magnetization and neutron diffraction suggest a simpler ferrimagnetic order ($T_{C}$${\sim}$49 K) between Mn$^{2+}$ and Cr$^{3+}$ spins, the exchange interactions are more complex than that expected from a two sublattice magnetic system.
The separations between the on-site energies of the $d$-orbitals of Mn(1), Mn(2) and Cr obtained from Wannier function analysis are in agreement with the expected crystal field splittings.
The electronic structure calculations yield a ferrimagnetic insulating ground state even in absence of Hubbard $U$ which persists for a wide range of $U$.
\newline
Due to the ${\sim}$90$^{o}$ superexchange, nearest neighbour exchange constants obtained from DFT are an order of magnitude smaller than those reported for various magnetic perovskite and double-perovskite compounds. 
%
%
%
The Mn(1)-O-Mn(2) (out of plane and in-plane), Mn(1)-O-Cr and Mn(2)-O-Cr superexchange interactions are found to be anti-ferromagnetic, while the Cr-O-O-Cr super-superexchange is found to be ferromagnetic.
The Mn(2)-O-Cr superexchange is weaker than the Mn(1)-O-Cr super-exchange, thus effectively resulting in ferrimagnetism.
From a simple 3-site Hubbard model, we derived expressions for the antiferromagnetic superexchange strength $J_{AFM}$ and the weaker ferromagnetic $J_{FM}$. 
%
%
The relative strengths of $J_{AFM}$ for the various superexchange interactions are in agreement with those obtained from DFT. 
The expression for Cr-O-O-Cr super-superexchange strength (${\tilde J}_{SS}$), which is derived considering a 4-site Hubbard model, predicts a ferromagnetic exchange in agreement with DFT.
%
Finally, our mean field calculations reveal that assuming a set of four magnetic sub-lattice for Mn$^{2+}$ spins and a single magnetic sublattice for Cr$^{3+}$ spins yields a much improved $T_{C}$, while a simple two magnetic sublattice model yields a much higher $T_{C}$.
\end{abstract}
%
%
%
%
%
\section{Introduction}
The perovskite oxide materials are known to exhibit a wide range of properties like complex magnetic structures, multiferroicity, metal-insulator transition and colossal magnetoresistance.
 The basic perovskite ($A$$B$O$_{3}$ : $A$-divalent element, $B$-transition metal (TM)) has a cubic structure (space group:$Pm{\bar 3}m$) e.g. SrCrO$_{3}$, SrTiO$_{3}$ \cite{KWLee_PRB2009, Gupta_PRB2004}. 
When $A$ is a trivalent ion (e.g. Rare earth or Y), structural distortions occur which include tilting of the $B$O$_{6}$ octahedra and the Jahn-Teller effect, due to which the crystal structure becomes orthorhombic.
Depending on the choice of $B$, diverse properties are observed, e.g. multiferroicity, orbital and charge ordering in manganites, high spin to low spin transition in cobaltates, spin reorientation in orthoferrites and orthochromates\cite{Malashevich_PRL2008, Goff_PRB2004, Pandey_PRB2008, Cao_2014, Yamaguchi_1974}.
\newline
 Addition of another transition metal atom at $B$-site in a regular pattern, in the cubic perovskite unit cell results in doubling of the unit cell. The new compound, $A_{2}$$BB'O_{6}$ with the space group $Fm{\bar 3}m$, is known as the double perovskite. Most of the double perovskite compounds have a rock-salt arrangement of the $B/B'$ atoms. 
Some examples are Ba$_{2}$MnWO$_{6}$, Sr$_{2}$MnWO$_{6}$, Ba$_{2}$PrIrO$_{6}$ and Ba$_{2}$YIrO$_{6}$\cite{Andrews2015, Fu2005, Dey_PRB2016}. 
The cubic double-perovskite structure is unstable for most of the elements at the $B/B'$ sites. Increase in the difference in the size of the $A$- and $B$-site cations reduces the tolerance factor below unity, thereby resulting in tetragonal ($I4/m$)\cite{Ayala_JAP2007}, ($P4_{2}/n$)\cite{Fujioka_JPhysChem2006}, rhombohedral($R{\bar 3}$)\cite{Fu_JALCOM2005}, and the most common monoclinic space group ($P2_{1}/n$)\cite{Anderson_1993, Cao_PRB2001} structure. 
An example of the double perovskites with monoclinic structure is the well-studied $R_{2}$MnNiO$_{6}$ series\cite{Booth_MRB2009}. All the members of this series are ferromagnetic insulators with a wide range of Curie temperatures from 80 to 280\,K. The ferromagnetic insulating nature for these is also predicted through density functional theory calculations. 
When $A$ is a divalent element, $B$/$B'$ belong to $3d$ and $4d$/$5d$ groups, respectively, the compounds show ferrimagnetism with a Curie temperature above 300\,K, and also a large magnetoresistance\cite{Jeng_PRB2003, Sanyal_PRB2016, Kobayashi_nature1998}. 
\newline
When $B'$ is a non-magnetic $p$-element (e.g.: Sb), the magnetism arises entirely due to the $B$-sublattice. Due to a greater separation between the $B$-site atoms. there occurs a drastic reduction in the magnetic transition temperature along with possible occurance of magnetic frustrations \cite{Retuerto2006, Battle1995}. 
For instance, Ca$_{2}$CrSbO$_{6}$ is a ferromagnet with a $T_{C}$ of 16\,K \cite{Retuerto2006}, while Sr$_{2}$CrSbO$_{6}$ is an antiferromagnet with a $T_{N}$ value of 12\,K \cite{Retuerto2006, Retuerto2007_JMatchem}, eventhough both the compounds crystallize in the monoclinic ($P2_{1}/n$) structure.
%
Density functional theory studies on Ca$_{2}$CrSbO$_{6}$ reveal the role of greater structural distortion in stabilizing the ferromagnetic ground state\cite{Baidya_PRB2012}.
\newline
Recently, double perovskites of the form Mn$_{2}$$M$SbO$_{6}$ and Mn$_{2}$$M$ReO$_{6}$ ($M$: Sc, Cr, Fe, Co and Mn)\cite{Antonio_Dalton2015, Antonio_JPCM2015b, Solana_ChemComm2019} have been studied experimentally. These compounds also have a monoclinic ($P2_{1}/n$) structure with the Mn atom at the $A$-site in the 2+ ionic state and  a four-fold oxygen coordination. 
%
Combining Mn and Ca results in the solid solution Ca$_{2-x}$Mn$_{x}$$M$SbO$_{6}$ which retains the single phase monoclinic structure for 0$<$$x$$<$0.17 and 1.73$<$$x$$<$2. For 0.74$<$$x$$<$1.1, the compositions crystallize in the tetragonal $P4_{2}/n$ structure while the remaining composition ranges show a two-phase co-existence\cite{McNally_PRM2020}. 
We have chosen the composition with $x$$=$1, also known as the double-double perovskite compound, with the general chemical formula $AA'BB'O_{6}$, as the subject of the present study. These compounds show a dual cationic order, with a columnar arrangement of the $A/A'$ atoms and a rocksalt arrangement of the $B/B'$ atoms.
The lattice parameters of these compounds are close to that of the cubic $Fm{\bar 3}m$ unit cell\cite{McNally_PRM2020}.
Similar to Mn$_{2}$$M$SbO$_{6}$, double-double perovskites also require high pressures for synthesis. 
\newline
The present work is a density functional theory (DFT) based study on a recently synthesized double-double perovskite CaMnCrSbO$_{6}$\cite{Madruga_PRM2021}, belonging to the Ca$_{2-x}$Mn$_{x}$CrSbO$_{6}$ series. The two end compounds of this series, Ca$_{2}$CrSbO$_{6}$ and Mn$_{2}$CrSbO$_{6}$ have a monoclinic structure ($P2_{1}/n$), but show entirely different magnetic behaviour. 
The former, shows a long range ferromagnetic order\cite{Retuerto2006}
%
while Mn$_{2}$CrSbO$_{6}$ orders antiferromagnetically at $T_{N}$$=$55\,K, though with a more complex propagation vector ${\vec k}$=(0.5,0,0.5)\cite{Antonio_Dalton2015}.
Thus with the presence of two different magnetic ions in the unit cell, one could expect that the multiple exchange interactions may lead to a complex long range magnetic structure or spin glass like behaviour in CaMnCrSbO$_{6}$.
Additionally, with 50${\%}$ replacement of Mn by Ca, and a larger separation between the Cr atoms, one may also expect a considerable reduction in the transition temperature.
However, magnetic neutron diffraction and bulk magnetization studies on CaMnCrSbO$_{6}$ confirm a collinear ferrimagnetic structure with oppositely aligned Mn$^{2+}$ and Cr$^{3+}$ spins having an ordering temperature, $T_{C}$$=$49 K, which is close to that of Mn$_{2}$CrSbO$_{6}$. 
 The structural neutron diffraction studies also show that due to similar ionic radii, anti-site disorder occurs between the Cr/Mn and Cr/Sb sites, which affects the magnetic properties of the system \cite{Madruga_PRM2021}. The different ligand symmetries of two non-equivalent Mn atoms in the unit cell could result in exchange interactions of different magnitudes between Mn and Cr, thus affecting the transition temperature along with the nature of magnetic order.
Thus, to understand the magnetic structure in the structurally ``ordered" unit cell, it is essential to estimate the magnitude and signs of various exchange constants ($J$) present in the system, which forms an important part of the present work. 
The origins for these exchange interactions are analyzed in terms of hopping between the different metal-and ligand-Wannier orbitals. 
Using a simpler formalism of a 3-site Hubbard model with a ground state comprising of single occupied $d$-orbial at each transition metal ion site and a doubly occupied O $p$-orbital, we have estimated the Mn-O-Mn, Mn-O-Cr superexchange interaction strengths. 
Similarly, the more complex Cr-O-O-Cr super-superexchange strength is also estimated from a 4-site Hubbard model with similar ground state as above.
In order to determine $T_{C}$, we consider a model with four Mn$^{2+}$ magnetic sub-lattices and one magnetic sub-lattice for the Cr$^{3+}$ spins. 
The results thus obtained provide good agreement with the experimental observations and offer a comprehensive picture of the magnetic structure of CaMnCrSbO$_{6}$.
\newline
This paper is organized as follows: after presenting the computational details in Section 2, we discuss in Section 3, the crystal structure and the local point group symmetry of the two non-equivalent Mn atoms and of the Cr atom in the unit cell.
Section 4.1 comprises of the non-magnetic density of states along with the Wannier function analysis. In Section 4.2, we discuss the spin-polarized electronic structure, effect of Hubbard $U$ and the magnetic structures along with the anisotropy of the system. The effects of anti-site disorder on the electronic structure and magnetism are discussed in Section 4.3. 
We have extracted the various exchange interactions in Section 4.4, while Section 4.5 comprises of discussion on various superexchange mechanisms in terms of hopping between various metal-ligand hopping integrals, with expression for $J$ derived using 3-site and 4-site Hubbard model.
Finally in Section 4.6, we present a detailed calculation on estimation of the magnetic transition temperature using the mean field theory.
%
\section{Computational details}
The calculations have been performed using DFT as implemented in the Vienna ${\it ab}$ ${\it initio}$ simulation package (VASP) which uses the projector augmented wave (PAW) method\cite{GKresse_PRB1996}. 
The Perdew-Burke-Ernzerhof (PBE) potential has been employed within the GGA (Generalized Gradient Approximation) and GGA+$U$\cite{JPPerdew_PRL1996, VIAnisimov_PRB1993}. 
A cut-off energy of 500 eV is used in the expansion of the plane waves. A 9$\times$9$\times$9 Monkhorst-Pack $k$-mesh centered at ${\Gamma}$ point in Brillouin zone (BZ) is used for performing the BZ integrations. 
%
%
The electrons from Cr/Mn: $3d$, $4s$, O: $2s$, $2p$ and Sb: $4f$, $5p, 5d$, $6s$ and Ca: $3p$, $4s$ states were considered as the valence electrons. Non-magnetic as well as spin-polarized self-consistent calculations have been performed. In addition, non-collinear calculations were also performed within the GGA+U+SO formalism.
\newline
Further, to obtain the local orbital representation, real-space Wannier functions derived from Mn and Cr $3d$ orbitals are constructed in the non-magnetic (NM) phase. Disentanglement has been achieved by employing a suitable energy window around the Fermi energy.
The crystal-field-split onsite energies and hopping amplitudes have also been derived. 
\section{Crystal Structure and local symmetry}
\begin{figure*}
 \centering
 \includegraphics[scale=0.5]{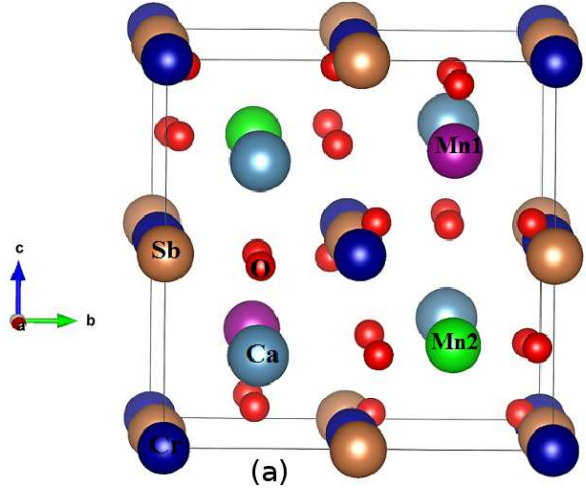}
\caption{Crystal structure of CaMnCrSbO$_{6}$. The inequivalent Mn atoms are marked as Mn1 and Mn2.}
\label{crystal_structure}
\end{figure*}
The double-double perovskite compound CaMnCrSbO$_{6}$ crystallizes in the tetragonal space group $P4_{2}n$. The unit cell (Fig.~\ref{crystal_structure}) comprises of four formula units: 24 O, 4 Mn, 4 Cr, 4 Sb and 4 Ca atoms. The site symmetries of the atoms are as follows: Ca - $4e$, O - $8g$, Sb - $4c$, Cr - $4c$, Mn(1) - $2a$ and Mn(2) - $2b$, respectively. The Mn(1) atoms are coordinated by four oxygen atoms in tetrahedral symmetry (Fig.~\ref{dosnm_cef}d), while the Mn(2) atoms are coordinated by four oxygen atoms in a square-planar geometry (Fig.~\ref{dosnm_cef}e). The Cr atoms occupy the $4c$-site with a six-fold coordination of oxygen atoms in a highly distorted octahedra, showing both, orthorhombic and GdFeO$_{3}$-like distortions(Fig.~\ref{dosnm_cef}f). In this structural arrangement, the Mn(1), Mn(2) and Cr ions have formal valency of +2, +2 and +3, respectively. 
\newline
\section{Results and discussion}
\subsection{Non-magnetic electronic structure, Wannier orbitals and on-site energies}
\begin{figure}
\centering
\minipage{0.5\textwidth}
 \includegraphics[width=\linewidth]{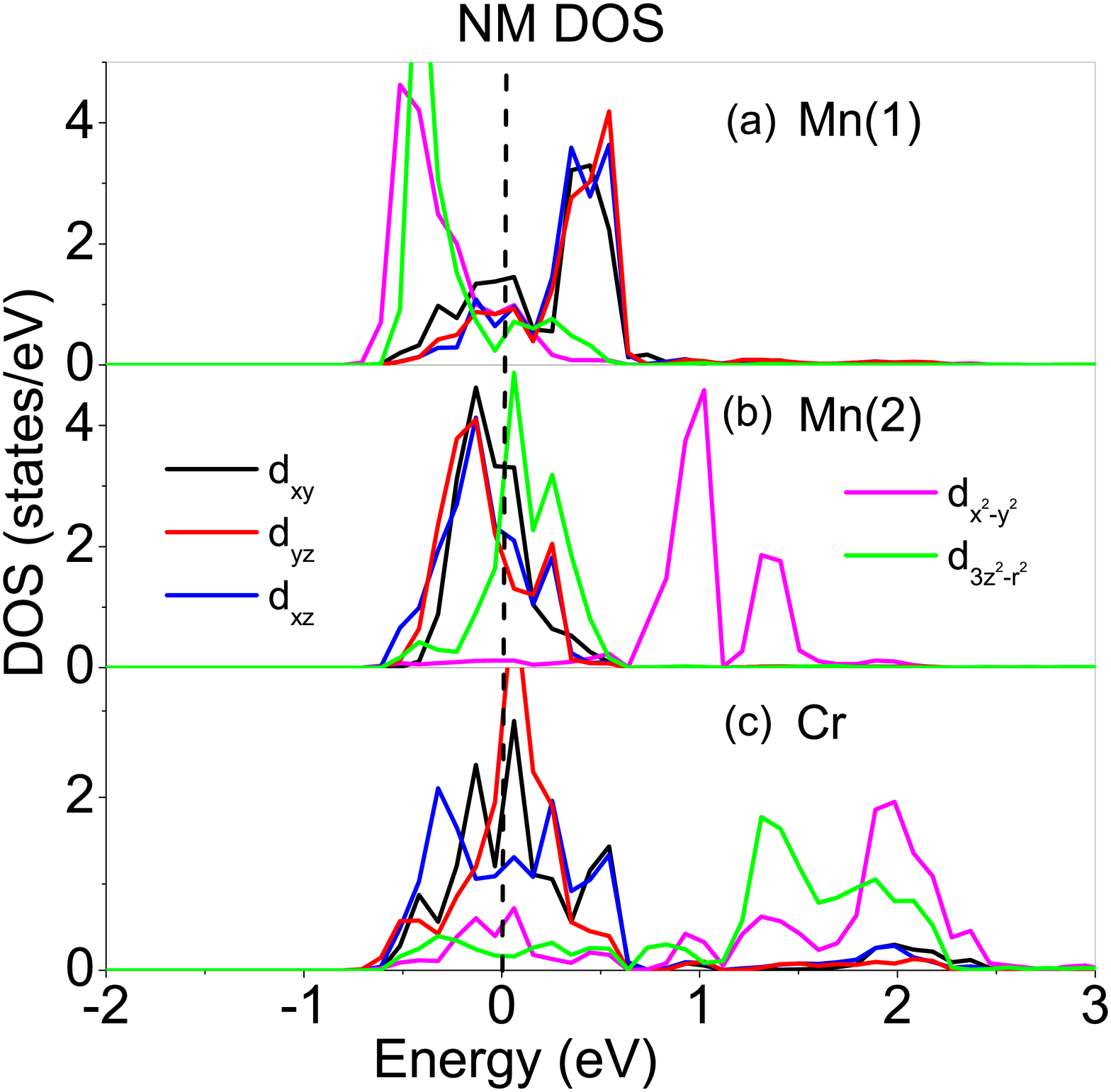}
\endminipage\hfill
\minipage{0.5\textwidth}
 \includegraphics[width=\linewidth]{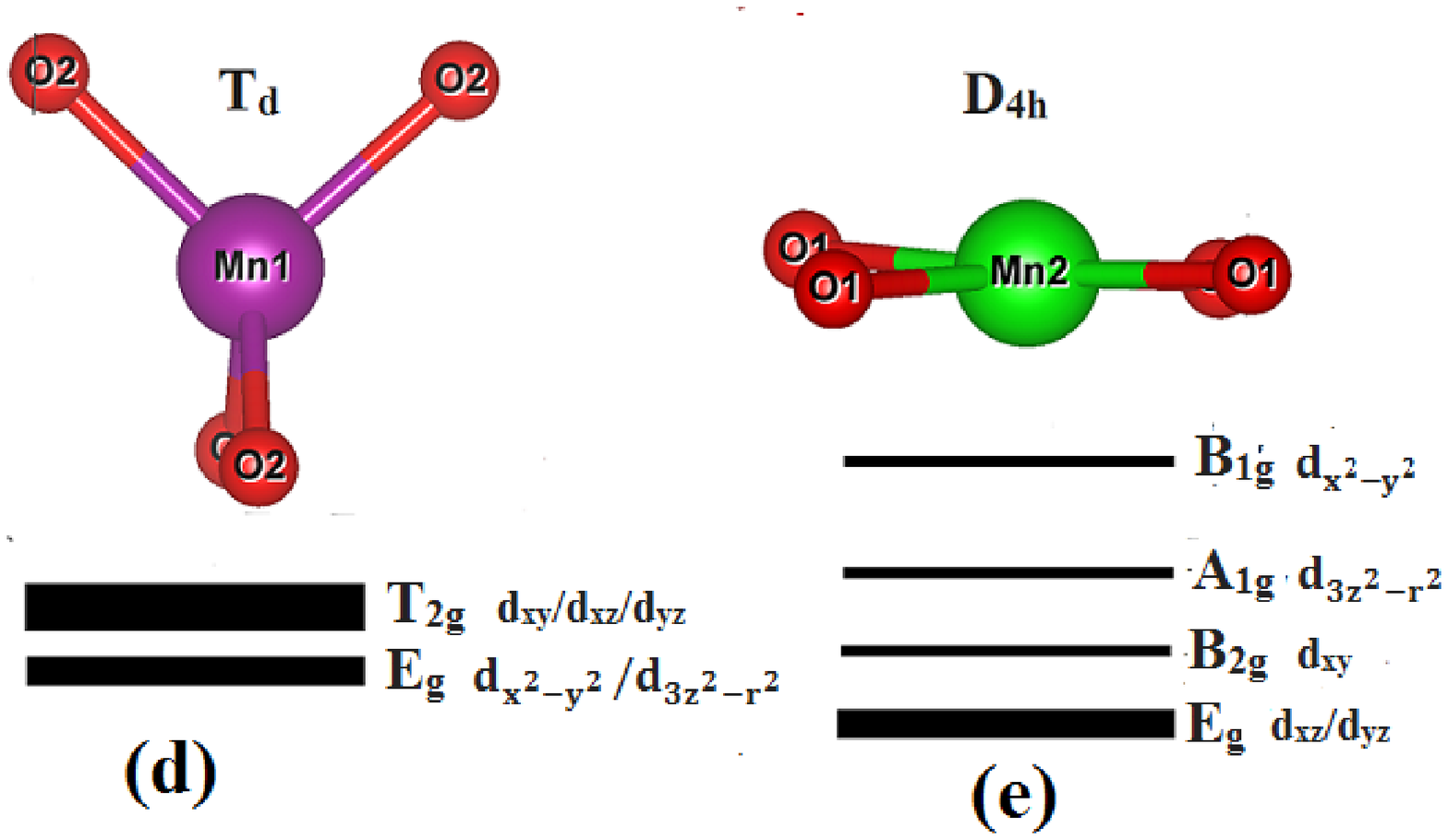}
\endminipage\hfill
\minipage{0.5\textwidth}
 \includegraphics[width=\linewidth]{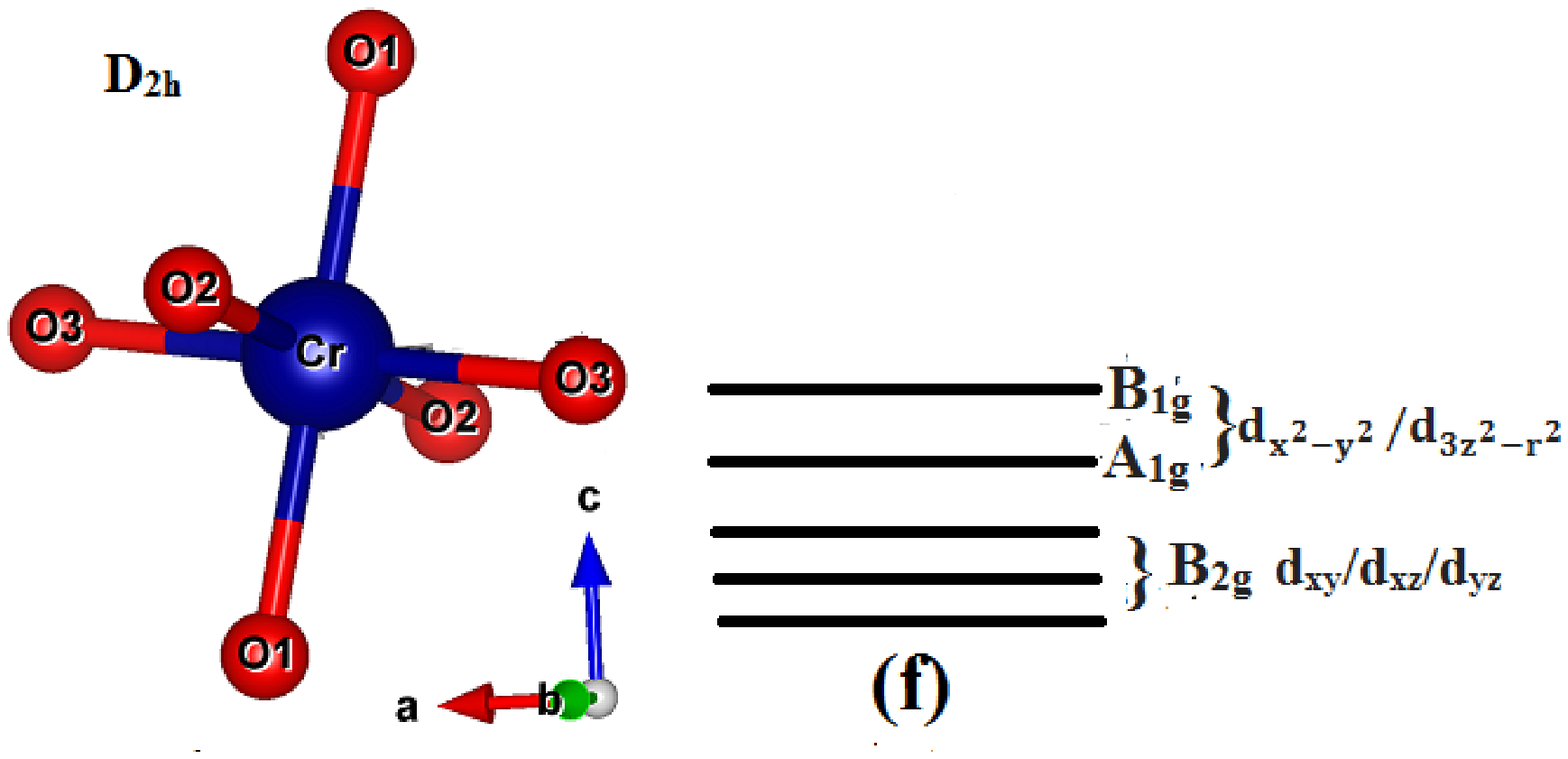}
\endminipage\hfill
\caption{(a)-(c) Site-resolved non-magnetic partial $d$ DOS of CaMnCrSbO$_{6}$. The Fermi energy is at 0 eV. (d)-(f) The local point group symmetries along with corresponding crystal field splitting schemes for (d) Mn(1), (e) Mn(2) and (f) Cr.} 
\label{dosnm_cef}
\end{figure}
The non-magnetic electronic structure calculations yield a metallic state for CaMnCrSbO$_{6}$ as seen from the density of states (DOS) (Figs.~\ref{dosnm_cef}(a-c)). All the five $d$ orbitals show significant spectral weight at the Fermi energy ($E_{F}$), while the O $p$ orbitals  have a smaller spectral weight in the DOS (not shown here).
The effect of the crystal-field splittings on the different magnetic ions is seen in the partial DOS.  
Due to tetrahedral crystal field, the $d$-orbitals of Mn(1)$^{2+}$ ion split into a two-fold degenerate $E_{g}$ ($d_{x^{2}-y^{2}}$ and $d_{3z^{2}-r^{2}}$) orbitals which are at lower energy, while the three-fold degenerate $T_{2g}$($d_{xy}$, $d_{xz}$ and $d_{yz}$) orbitals are at higher energy(Fig.~\ref{dosnm_cef}d). This splitting is in reverse order with respect to the usual $d$-orbital splitting in cubic crystalline fields.
Fig.~\ref{dosnm_cef}a, shows a clear separation between the $E_{g}$ and $T_{2g}$ DOS of the Mn(1) atom with peaks at ${\sim}$-0.5 and +0.5 eV respectively in agreement with the local symmetry (Fig.~\ref{dosnm_cef}d). 
\newline
The $d$-orbitals of the Mn(2)$^{2+}$ ion, which experience a planar crystal field, split into a set of four levels. The lowest level is made of the two-fold degenerate $E_{g}$ comprising of $d_{xz}$/$d_{yz}$ orbitals. The $d_{xy}$ orbital, belonging now to the $B_{2g}$ representation, lies above the $d_{xz}$/$d_{yz}$ orbitals (Fig.~\ref{dosnm_cef}e). The $d_{x^{2}-y^{2}}$ and $d_{3z^{2}-r^{2}}$ orbitals are non-degenerate and belong to the $B_{1g}$ and $A_{1g}$ representations, respectively. The $d_{x^{2}-y^{2}}$ orbital has the highest energy.
The $d$ DOS of Mn(2) atom (Fig.~\ref{dosnm_cef}b) shows features in accordance with the planar geometry. 
With almost identical spectral features, the $d_{xy}$ and $d_{xz}$/$d_{yz}$ DOS of the Mn(2) atom are close in energy. The DOS of all the three states pass through $E_{F}$ with a peak at -0.2 eV. While the $d_{3z^{2}-r^{2}}$ DOS shows a large peak at $E_{F}$, the $d_{x^{2}-y^{2}}$ corresponds to the highest energy level with a peak in DOS at +1 eV. 
\newline
For the Cr$^{3+}$ ion, being at the centre of an orthorhombically distorted octahedron, the $T_{2g}$ ($d_{xz}$, $d_{yz}$ and $d_{xy}$) and $E_{g}$ ($d_{x^{2}-y^{2}}$ and $d_{3z^{2}-r^{2}}$) orbitals undergo additional splitting, lifting the degeneracy completely. As shown in Fig.~\ref{dosnm_cef}f, the $T_{2g}$ orbitals of Cr$^{3+}$ split to form a set of three non-degenerate $B_{2g}$ orbitals, while the degenerate $E_{g}$ orbitals form a linear combination of $d_{x^{2}-y^{2}}$ and $d_{3z^{2}-r^{2}}$ orbitals resulting in two non-degenerate $A_{1g}$ and $B_{1g}$ orbitals.
The $d$ DOS of Cr (Fig.~\ref{dosnm_cef}c) comprises of partially occupied $d_{xy}$, $d_{xz}$ and $d_{yz}$ states at $E_{F}$, while the $d_{x^{2}-y^{2}}$ and $d_{3z^{2}-r^{2}}$ states occur mostly above $E_{F}$, with broad peaks around +1.2 and +2.0 eV, respectively. Thus these are in agreement with the expected crystal-field splitting scheme.
%
\begin{figure}
\centering
\minipage{1.05\textwidth}
 \includegraphics[width=\linewidth]{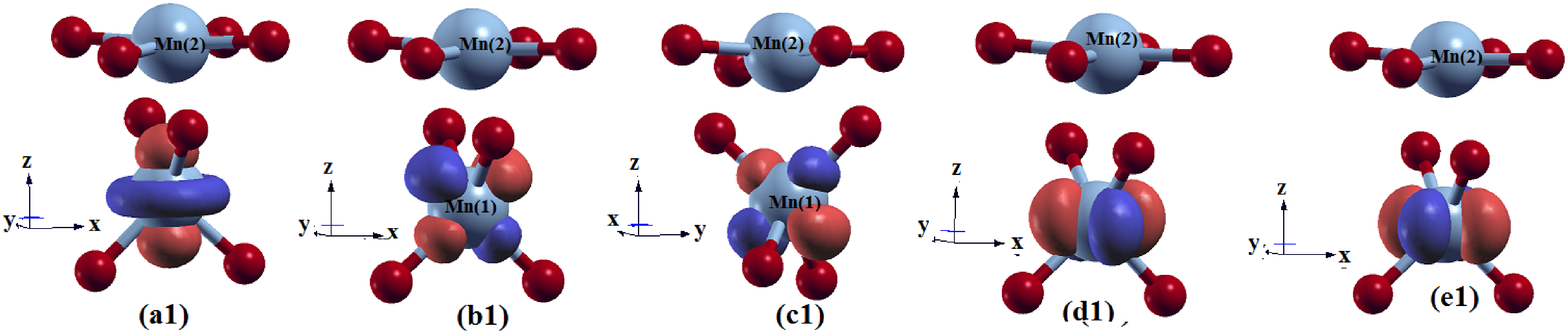}
\endminipage\hfill
\minipage{1.05\textwidth}
 \includegraphics[width=\linewidth]{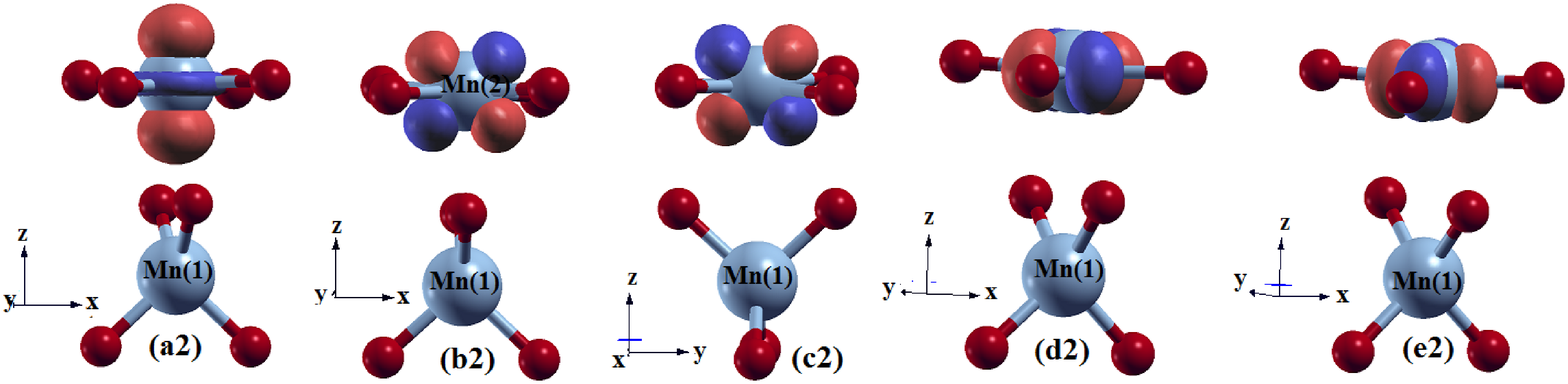}
\endminipage\hfill
\minipage{1.05\textwidth}
 \includegraphics[width=\linewidth]{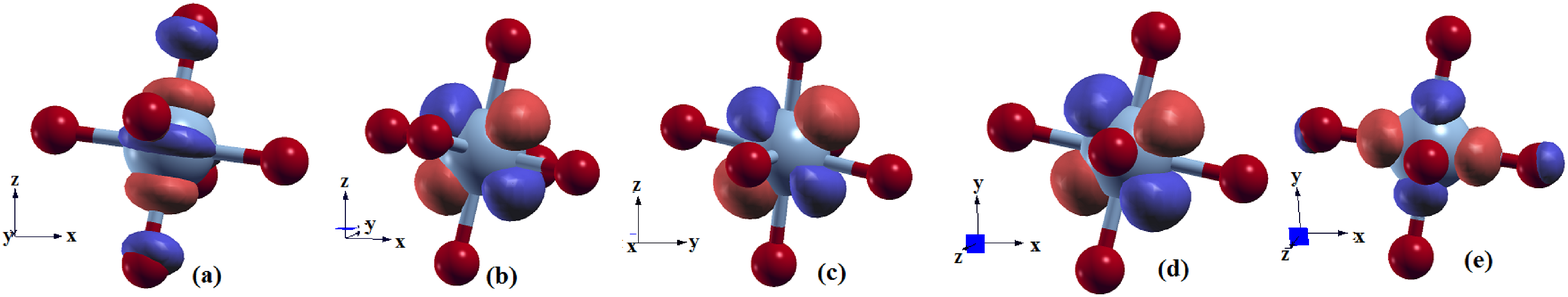}
\endminipage\hfill
\caption{(a1)-(e1), (a2)-(e2) and (a)-(e) correspond to the $d_{3z^{2}-r^{2}}$, $d_{xz}$, $d_{yz}$, $d_{xy}$ and $d_{x^{2}-y^{2}}$ Wannier $d$- orbitals of Mn(1), Mn(2) and Cr, respectively, in CaMnCrSbO$_{6}$. The $x$, $y$ and $z$ axes correspond to the $a$, $b$ and $c$ axes of the crystal.} 
\label{Wannier}
\end{figure}
\newline
In order to estimate the crystal-field splitting in the transition metal (TM) ions, along with the various hopping integrals, the Bloch states obtained in terms of augmented plane wave basis have been expressed in terms of real space Wannier orbitals, considering the $d$-orbitals of both, Mn and Cr along with the $p$-orbitals of oxygen.
In Fig.~\ref{Wannier}, we show the Wannier orbitals of Mn(1), Mn(2) and Cr atoms. 
For the Mn(1) atom (Fig.~\ref{Wannier}(a1-e1) the $z$-axis is along the crystallographic $c$-direction, while the $x$- and $y$-axes are diagonal to the $a$ and $b$-axes of the crystal. In case of the Mn(2) atom, (Fig.~\ref{Wannier}(a2-e2)), the $x$- and $y$-axes are along the lines joining the centres of the Mn(2) and planar O atoms, while the $z$-axis, similar to the case of Mn(1) is along the crystallographic $c$-direction. The local coordinate axes for the Cr atom are along the lines joining the centres of the the Cr and O atoms of the octahedron and thus tilted w.r.t. the global crystallographic axes (Fig.~\ref{Wannier}(a-e)).
\newline
From the Wannier analysis, we obtain the real-space matrix elements of the Hamiltonian, i.e. the hopping amplitudes between the TM ion and oxygen atoms. The diagonal elements, which correspond to the the on-site energies of Mn(1), Mn(2) and Cr $d$ states, are listed in Table 1.
The on-site energies of the $d$ states in Mn(1), Mn(2) and Cr follow the crystal-field splitting scheme as is shown in Fig.~\ref{dosnm_cef}(d-f).It should be mentioned that in the Mn(2) atom, the on-site energy of the $d_{xy}$ state obtained is slightly lower than $d_{xz}$/$d_{yz}$ states by 0.06 eV, unlike as shown in Fig.~\ref{dosnm_cef}e.
The total crystal-field splitting is the smallest for Mn(1), as is expected for an Mn(1)$^{2+}$ ion placed in a tetrahedral environment, while the Cr$^{3+}$ ion placed in a highly distorted octahedral environment exhibits the maximum splitting. 
These hopping amplitudes would be utilized later, in the context of superexchange interactions.
\begin{table}
\caption{\label{tabone}On-site energies (eV) of the crystal field levels of Mn(1)$^{2+}$, Mn(2)$^{2+}$ and Cr$^{3+}$.} 
\begin{center}
\lineup
\item[]\begin{tabular}{@{}*{10}{l}}
\br                           
$\0\0$&$d_{xy}$  & $d_{xz}$ & $d_{yz}$ & $d_{x^{2}-y^{2}}$ & $d_{3z^{2}-r^{2}}$ & $E_{max}$$-$$E_{min}$ (eV)\cr
\mr
\0Mn(1)$^{2+}$&6.37&6.32 &6.31&6.14&6.03 &\m\m 0.34\cr
\0Mn(2)$^{2+}$&6.32&6.38&6.38&6.82&6.51 &\m\m 0.50\cr
\0Cr$^{3+}$&6.11&6.06&6.13&6.90&6.66 &\m\m 0.84\cr
\br
\end{tabular}
\end{center}
\end{table}
\subsection{Spin-polarized electronic structure and effect of correlations}
The spin polarized calculations have been carried out for the ferrimagnetic (FiM) configuration, with  
the Mn$^{2+}$ and Cr$^{3+}$ spins arranged as spin-up and spin-down, respectively.
 Additionally, the calculations have also been performed for ferromagnetic (FM) and various anti-ferromagnetic (AFM) arrangements, viz. A-type, C-type, and G-type. 
%
The calculations, for all the magnetic configurations, have been performed for different values of $U_{eff}$($=$$U$$-$$J_H$; $U$ is the Coulomb interaction energy of the $d$-orbitals, $J_{H}$ is the Hund's exchange energy). The $U$ value was varied between 0 and 6 eV, with $J_H$$=$1 eV for both, Mn and Cr. Calculations were also performed for unequal values of $U_{eff}$ in Mn and Cr.  
For AFM calculations, either the Mn or the Cr sublattice was fixed in FM configuration, while the other sublattice was arranged as A-type, C-type and G-type. 
As seen from Table 2, at $U_{eff}$$=$0 eV, the FiM configuration is prefered by a large margin, while the FM configuration has the highest energy. We note that with the increase in $U_{eff}$, the difference between the energies of various configurations gets reduced and at $U_{eff}$$=$5 eV, the FiM no longer corresponds to the lowest energy configuration.
\Table{\label{tabl2}Total energy per unit cell (in meV) w.r.t that for the FM configuration for various magnetic configurations and different $U_{eff}$$=$$U$-$J_{H}$ (in eV) values. The values of $U$ and $J$ are same for Mn and Cr, unless mentioned otherwise.}
\br
&&&\centre{3}{Relative energy (meV)}\\
\ns
 &&\crule{4}\\
Magnetic&&$U$$=$0&$U$$=$4 &$U$$=$6(Mn), 5.5(Cr) & $U$$=$6 \\
 Configuration            &&$J_H$$=$0&$J_H$$=$1 &$J_H$$=$1.5(Mn), 1(Cr) & $J_H$$=$1 \\
             && $U_{eff}$$=$0&$U_{eff}$$=$3 &$U_{eff}$$=$4.5(Mn, Cr) & $U_{eff}$$=$5 \\
\mr
FM   &    &      0.0           &    0.0          &\m\m\m\m      0.0        & 0.0 \\
        FiM  & &         \bf{-227.0}    &  \bf{-73.4}  &\m\m\m\m    -\bf{20.8}    &-10.8\\
Cr-A(AFM)   &     &     -62.9    &  -6.0  &\m\m\m\m 13.9       & 17.5\\
        Cr-C(AFM)   & &        -62.8    &  -3.5  &\m\m\m\m    16.6 &31.3 \\
        Cr-G(AFM)   & &        -62.8    &  -3.5  &\m\m\m\m 16.6    & 31.3\\
        Mn-A(AFM)   &  &        -127.2    &  -53.7  &\m\m\m\m      -16.5  & \bf{-15.4}\\
        Mn-C(AFM)   &  &       -131.9    &  -48.9  &\m\m\m\m -15.4   & -13.0\\
        Mn-G(AFM)   &   &      -94.5    &  -42.5  &\m\m\m\m -14.1    & -14.1 \\
\br
\end{tabular}
\end{indented}
\end{table}
\newline
\begin{figure}
\centering
\minipage{0.5\textwidth}
 \includegraphics[width=\linewidth]{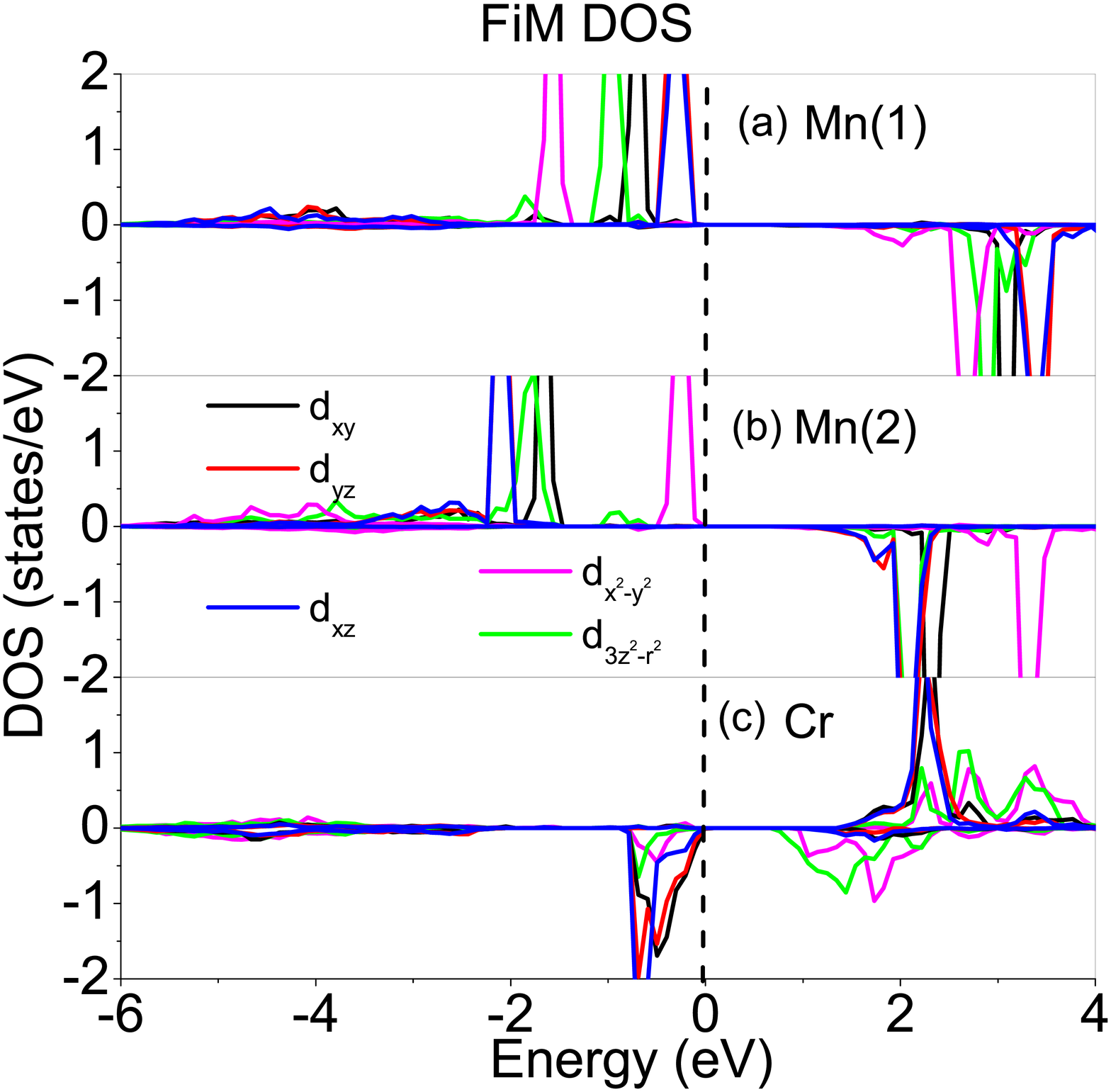}
\endminipage\hfill
\minipage{0.5\textwidth}
 \includegraphics[width=\linewidth]{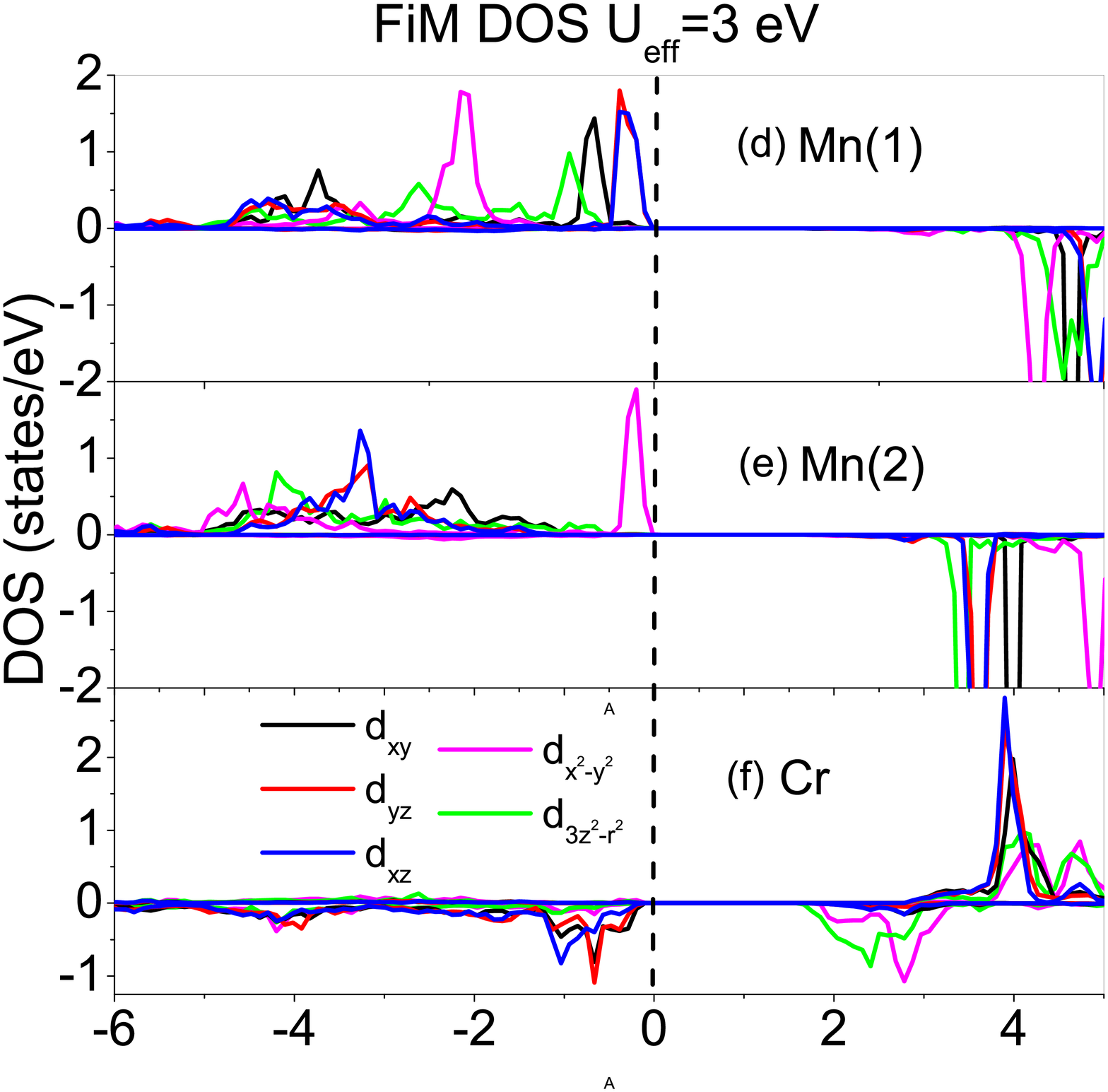}
\endminipage\hfill
\caption{Ferrimagnetic DOS of CaMnCrSbO$_{6}$ for $U_{eff}$=0 eV (left panel) and $U_{eff}$=3 eV.} 
\label{spinpolarized_dos}
\end{figure}
 In Figs.~\ref{spinpolarized_dos}(a-c), we show the spin polarized DOS of CaMnCrSbO$_{6}$ for the FiM configuration corresponding to $U_{eff}$$=$0 eV.
Even in the absence of Hubbard $U$, FiM CaMnCrSbO$_{6}$ shows an insulating behaviour with a fairly large band gap of ${\sim}$ 0.7 eV. 
As seen from Figs.~\ref{spinpolarized_dos}(a) and (b), the spin-up $d$ states of Mn(1) and Mn(2) are fully occupied, while spin-down $d$ states are completely empty. The energy scheme of the crystal field split levels in the NM DOS case (Fig.~\ref{dosnm_cef}) is found to be applicable in the FiM configuration as well. In the Mn(1) $d$-DOS (Fig.~\ref{spinpolarized_dos}a), the $d_{x^{2}-y^{2}}$ has the lowest energy, with a sharp peak at ${\sim}$ -1.8 eV. The $d_{xz}$/$d_{yz}$ states retain their near-degenerate character and are close to $E_{F}$, while the $d_{xy}$ gets shifted to -0.5 eV, preceded by the $d_{3z^{2}-r^{2}}$, which shows a peak at ${\sim}$ -1 eV.  
\newline
In Mn(2) spin-up DOS, the $d_{x^{2}-y^{2}}$ occurs in isolation near $E_{F}$ with a peak at ${\sim}$ -0.3 eV, while the spin-up DOS of the remaining four $d$ states shows sharp peaks between -1.7 and -2.0 eV. Above $E_{F}$, the peak for $d_{3z^{2}-r^{2}}$ along with $d_{xz}$ and $d_{yz}$ spin-down DOS occurs at ${\sim}$ +2.1 eV, while the peak for $d_{x^{2}-y^{2}}$ spin-down DOS occurs at ${\sim}$ +3.2 eV.
In Fig.~\ref{spinpolarized_dos}c, the spin-down states of Cr are partially occupied, with the $d_{xz}$ and $d_{yz}$ spin-down DOS showing sharp peaks at ${\sim}$ -0.5 eV, while the $d_{xy}$ DOS shows a broader feature, with a peak at ${\sim}$ -0.2 eV.
The $d_{3z^{2}-r^{2}}$ and $d_{x^{2}-y^{2}}$ spin-down DOS occur above $E_{F}$, overlapping with the unoccupied $T_{2g}$ spin-up DOS above +1.5 eV. 
\newline
In Figs.~\ref{spinpolarized_dos}(d-f), we show the FiM partial DOS of CaMnCrSbO$_{6}$ for $U_{eff}$$=$3 eV. The effect of $U$ is different for each atom, though there occurs a broadening, in general, for each of the DOS. For the Mn(1) atom, the separation between the spin-up $d_{x^{2}-y^{2}}$ and $d_{3z^{2}-r^{2}}$ DOS shows an increase, whereby the former now shows a peak at ${\sim}$ -2 eV. The spin-up $d_{xz}$/$d_{yz}$ states are nearest to $E_{F}$.
For Mn(2) atom, the spin-up $d_{x^{2}-y^{2}}$ DOS (Fig.~\ref{spinpolarized_dos}e ) retains its proximity to $E_{F}$. For the remaning $d$ states of Mn(2), the DOS is spread in the range -1 to -5 eV, with reduced spectral weight as compared to $U_{eff}$=0 eV case. However, the spectral width of the spin-down DOS remains uneffected, though the peaks are almost rigidly shifted by ${\sim}$1.5 eV.
The Cr spin-down DOS(Fig.~\ref{spinpolarized_dos}f) below $E_{F}$ partly retain their features in presence of the correlations, though with increased width and diminished intensity of peaks. 
The unoccupied Cr spin-down $d_{x^{2}-y^{2}}$ and $d_{3z^{2}-r^{2}}$ DOS are shifted to $~$+2 eV with accentuated peaks, thus resulting in a larger bandgap (${\sim}$ 1.7 eV) for CaMnCrSbO$_{6}$ with $U_{eff}$$=$3 eV.
\begin{figure*}[!htb]
  \centering
 \includegraphics[scale=0.20]{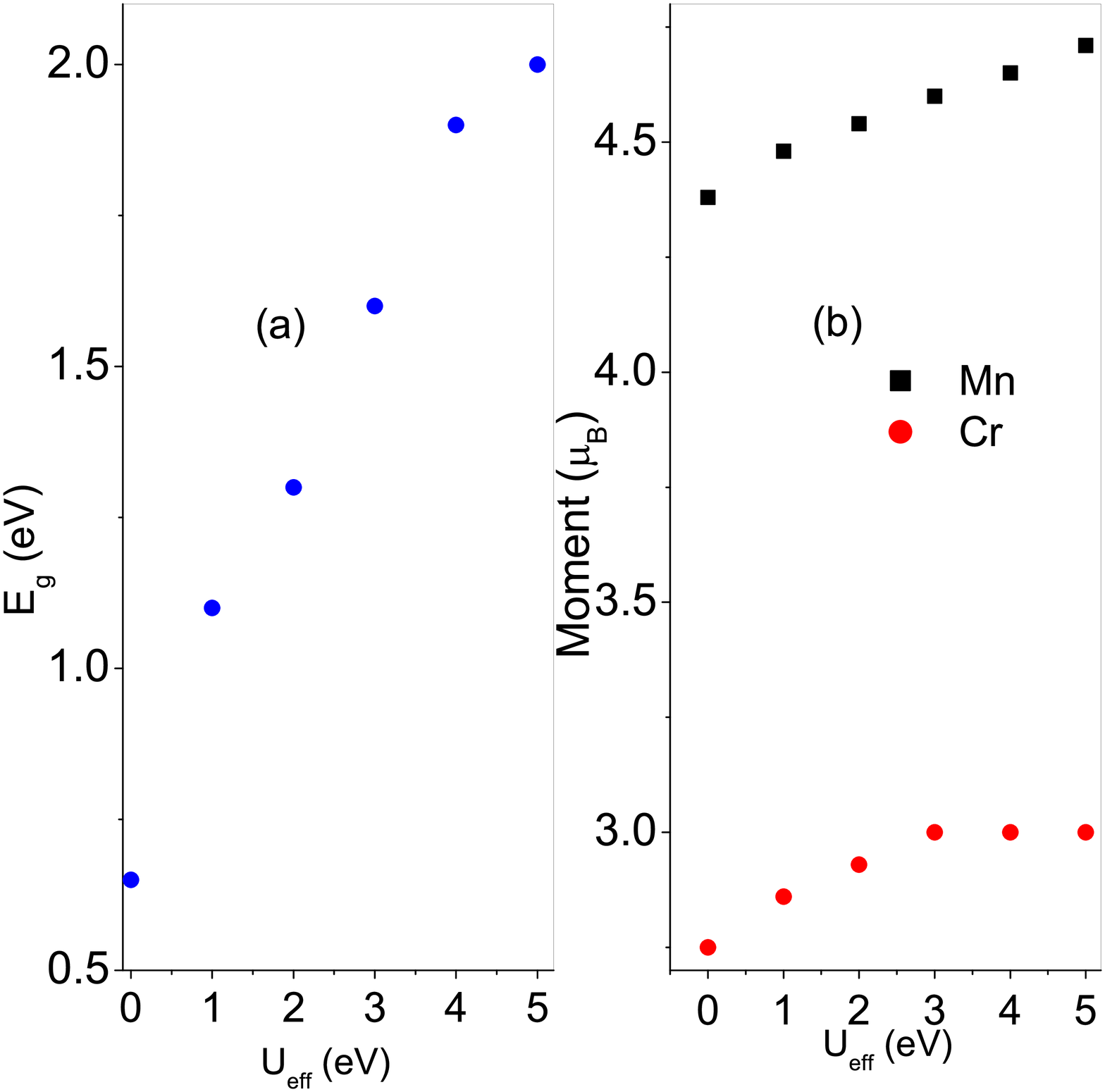}
\caption{Variation of (a) energy gap and (b) magnetic moments with $U_{eff}$. Mn(1) and Mn(2) show the same variation on this scale.}
\label{gap_moment_variation}
\end{figure*}
\newline
In Fig.~\ref{gap_moment_variation}(a), we show the variation of bandgap in CaMnCrSbO$_{6}$ as a function of $U_{eff}$.
There occurs a bandgap of ${\sim}$0.7 eV for $U_{eff}$$=$0 eV. After a sudden jump, with further increase in $U_{eff}$, the bandgap shows almost a linear variation with correlation energy. Thus CaMnCrSbO$_{6}$ shows the typical characteristics of a Mott-Hubbard insulator. 
In Fig.~\ref{gap_moment_variation}(b), we also show the variation of magnetic moment of Mn and Cr. The variation in moments of Mn(1) and Mn(2) are identical. While Cr attains the fully saturated moment of 3 ${\mu}_{B}$ for $U_{eff}$$=$3 eV, Mn(1) and Mn(2) show a slight reduction, from the expected value of 5 ${\mu}_{B}$ in their magnetic moments. 
\subsubsection{Non-collinear calculation and anisotropy}
Studies reveal that below the Neel temperature, CaMnCrSbO$_{6}$ orders magnetically in the ${\Gamma}_{3}$ representation\cite{Madruga_PRM2021}.
In this, the $x$ and $y$ components of the spins within the Mn(1)$^{2+}$ and Mn(2)$^{2+}$ sub-lattices are ferromagnetically coupled, though there is no restriction on the inter-sublattice coupling.
The four Cr$^{3+}$ spins in the unit cell are anti-ferromagnetically coupled in the $x$- and $z$-directions, and ferromagnetically coupled in the $y$-direction.
%
\newline
It may be noted that experimentally, it is not resolved, whether the spins are along the $x$ or the $y$-directions. 
To probe the anisotropy of the compound using DFT, non-collinear calculations have been performed by including the effects of spin-orbit coupling. We used $U$$=$6 eV and $J_{H}$$=$1.5 eV for Mn and $U$$=$5.5 eV and $J_{H}$$=$1 eV for Cr. Calculations were performed for various orientations of the spins. 
We find that the configuration with the Mn$^{3+}$ and Cr$^{3+}$ spins in the $x$-direction correspond to the lowest energy, while the spins in the $y$ and $z$-directions yield higher energy by 0.26 and 0.29 meV per unit cell respectively, thus conforming to the experimental results.
\subsection{Effect of anti-site disorders}
Structural studies in CaMnCrSbO$_{6}$ using neutron diffraction reveal that upto 20 ${\%}$ of the Mn(1) and Mn(2) sites are occupied by Cr atoms and vice versa, while upto 5 ${\%}$ of the Cr(Sb) sites are occupied by Sb(Cr) atoms, thus resulting in formation of anti-site disorders (ASD) \cite{Madruga_PRM2021}. 
The reduced magnetic moment of 3.8 ${\mu}_{B}$, obtained from magnetic neutron diffraction for both Mn$^{2+}$ spins, has been attributed to the ASDs \cite{Madruga_PRM2021}. We have made an effort to probe the effect of ASD on the magnetic and electronic structure of CaMnCrSbO$_{6}$. For this, we have adopted a simple picture and performed additional spin-polarized calculations with inclusion of ASDs.
The full-fledged calculations incorporating the reported switching between atoms (${\sim}$ 20${\%}$, 20${\%}$ and 5${\%}$ for Mn(1), Mn(2), Sb, respectively, with Cr atoms) would be computationally expensive requiring a very large supercell. 
Hence, instead, we tried to assess the effect of ASD in a simple manner by carrying out three independent calculations (in the FiM configuration, with $U_{eff}$$=$5 eV for Mn and Cr): one, where there is no ASD (labelled as “Ordered”), other with an exchange between Mn and Cr (labelled as ASD1), and the third where Sb and Cr are exchanged (labelled as ASD2). 
A supercell with lattice parameters ($a$, $a$, $2c$) was employed in the FiM configuration was used for all the three cases.
In this supercell there are 8 atoms each of Mn, Cr and Sb. In ASD1, one Cr atom is switched with an Mn atom (we chose Mn(1) site for this), while in ASD2, one Cr atom is switched with an Sb atom. The resulting percentage disorders (12.5${\%}$ in ASD1 and ASD2, each) within the supercell are fair representatives of those from structural analysis.
\begin{table}
\caption{\label{tabone}Magnetic moments and band gap in supercells of FiM CaMnCrSbO$_{6}$ for $U_{eff}$$=$5 eV in ordered and ASD configurations. The prime corresponds to the position after switching of the atoms. The total energy w.r.t. ordered supercell is also mentioned.} 
\begin{indented}
\lineup
\item[]\begin{tabular}{@{}*{12}{l}}
\br                              
$\0\0$&&Atom(Position)&Magnetic &Band & Total Energy& & &\cr
$\0\0$&&&Moment(${\mu}_{B}$)&gap (eV)& \m (eV) &&&\cr
\mr
 \0\0& Ordered   &  Mn(1)(0.25, 0.25, 0.25) &\m  4.71  &   \cr
\0\0  &   &  Mn(2)(0.75, 0.75, 0.25) &\m4.73 &\m2.0 &\m 0  \cr
\0\0  &   &  Cr(0, 0, 0) &\m-3.10 & &  \cr
\0\0  &   &  Cr(0.5, 0.5, 0) &\m-3.10 & &  \cr
\0\0  &   &  Sb(0, 0, 0.5) &\m0.00 & &  \cr
\0&&&&&&&&&&& \cr
 \0\0 & ASD1 &Mn(2)(0.75, 0.75, 0.25) &\m 4.71  \cr
  \0\0  &   &  Mn'(0, 0, 0) &\m-3.96 &\m0.1 &\m +2.18  \cr
   \0\0  &   &  Cr(0.5, 0.5, 0.0) &\m-3.10 &  \cr
    \0\0  &   &  Cr'(0.25, 0.25, 0.25) &\m3.96  \cr
   \0\0  &   &  Sb(0, 0, 0.5) &\m0.00 & &  \cr
   \0&&&&&&&&&&& \cr
  \0\0 & ASD2 &Mn(1)(0.25, 0.25, 0.25) &\m 4.71 &     \cr
  \0\0  &   &  Mn(2)(0.75, 0.75, 0.25) &\m 4.73 &\m 0.7 & \m +2.31  \cr
   \0\0  &   &  Cr(0.5, 0.5, 0) &\m -3.10 &  \cr
    \0\0  &   &  Cr'(0, 0, 0.5) &\m -3.06 &  \cr
     \0\0  &   &  Sb'(0, 0, 0) &\m 0.00 &  \cr
\br
\end{tabular}
\end{indented}
\end{table}
\begin{figure}[!htb]
\minipage{0.33\textwidth}
 \includegraphics[width=\linewidth]{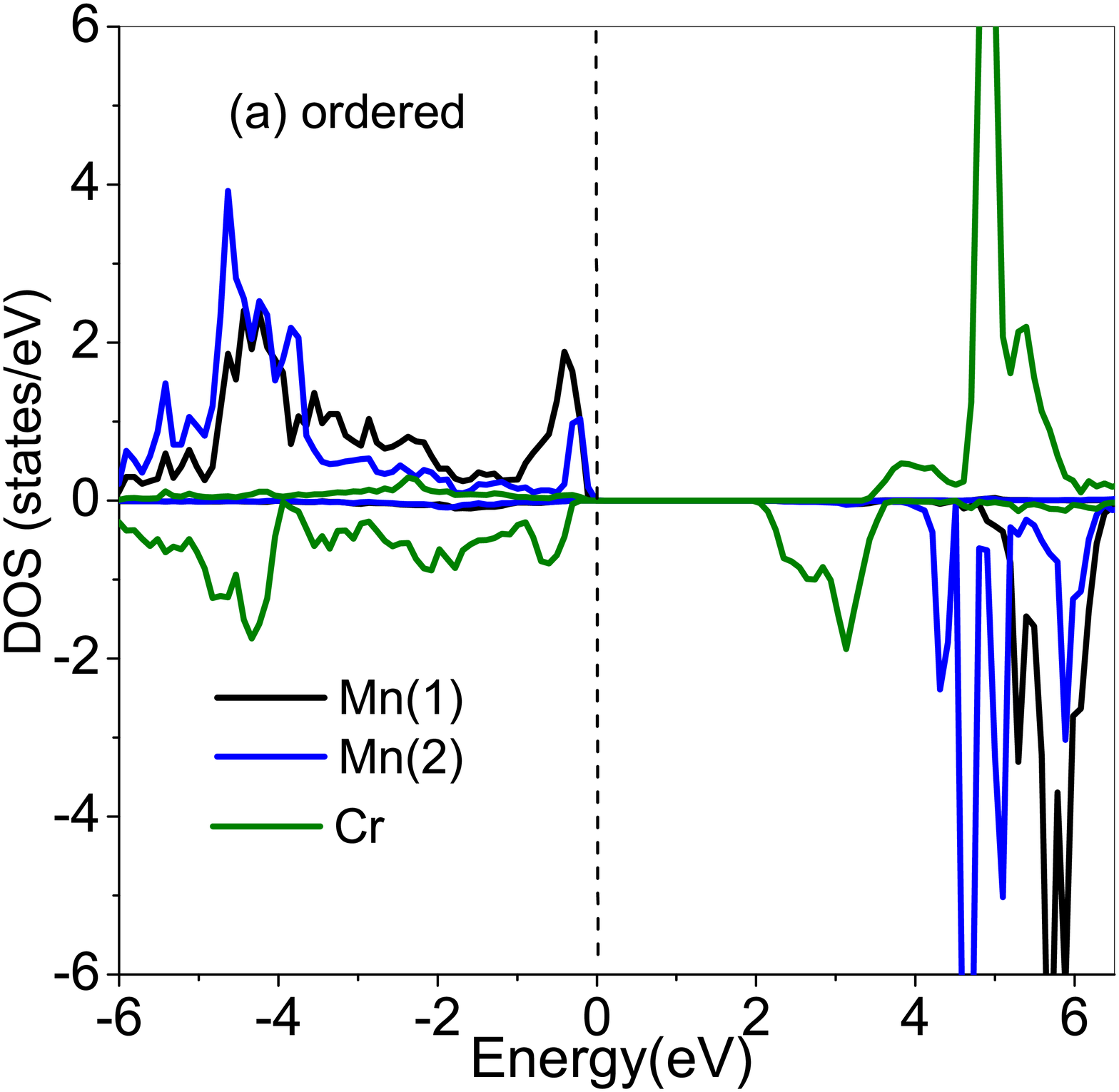}
\endminipage\hfill
\minipage{0.34\textwidth}
 \includegraphics[width=\linewidth]{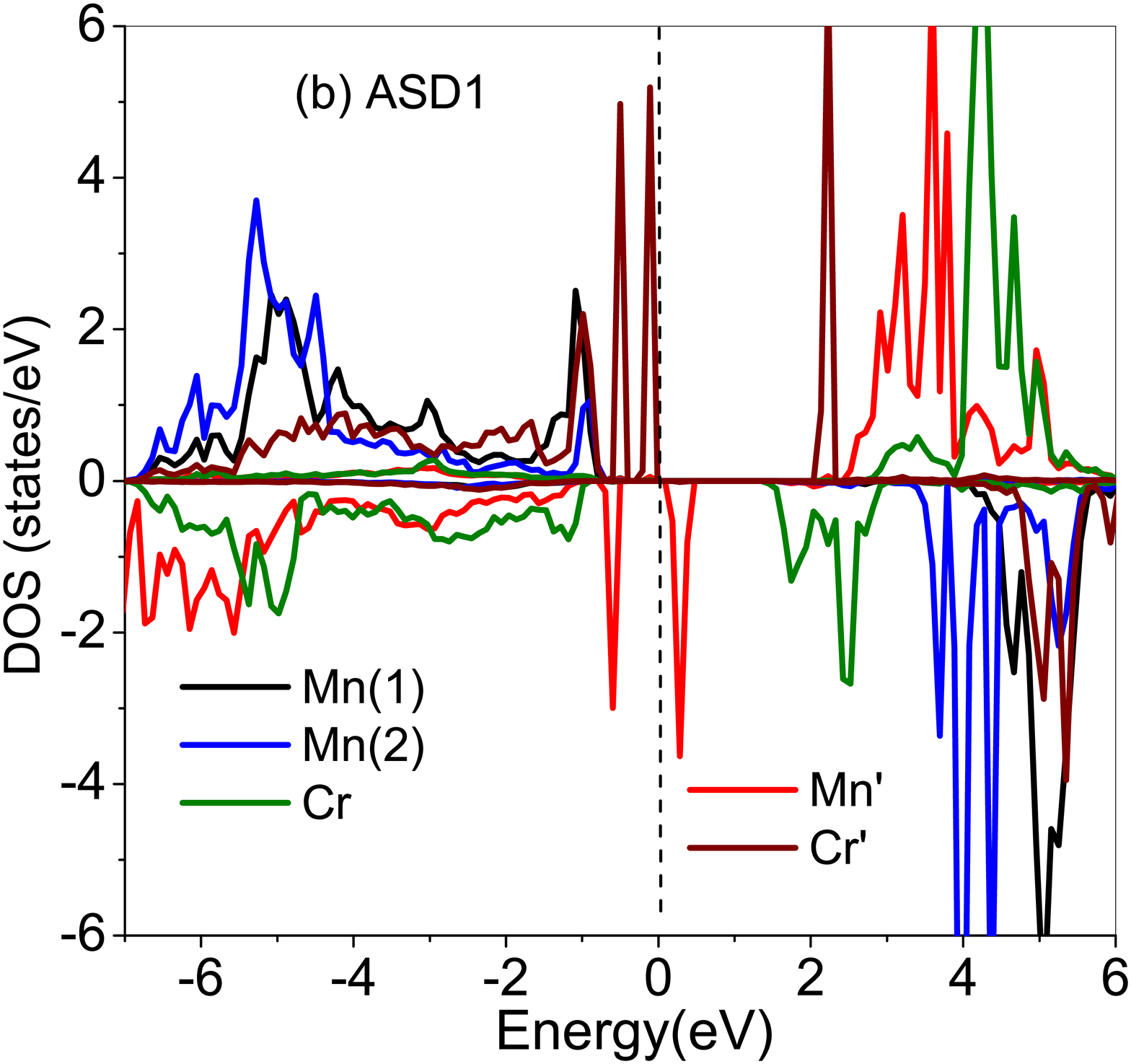}
\endminipage\hfill
\minipage{0.33\textwidth}
 \includegraphics[width=\linewidth]{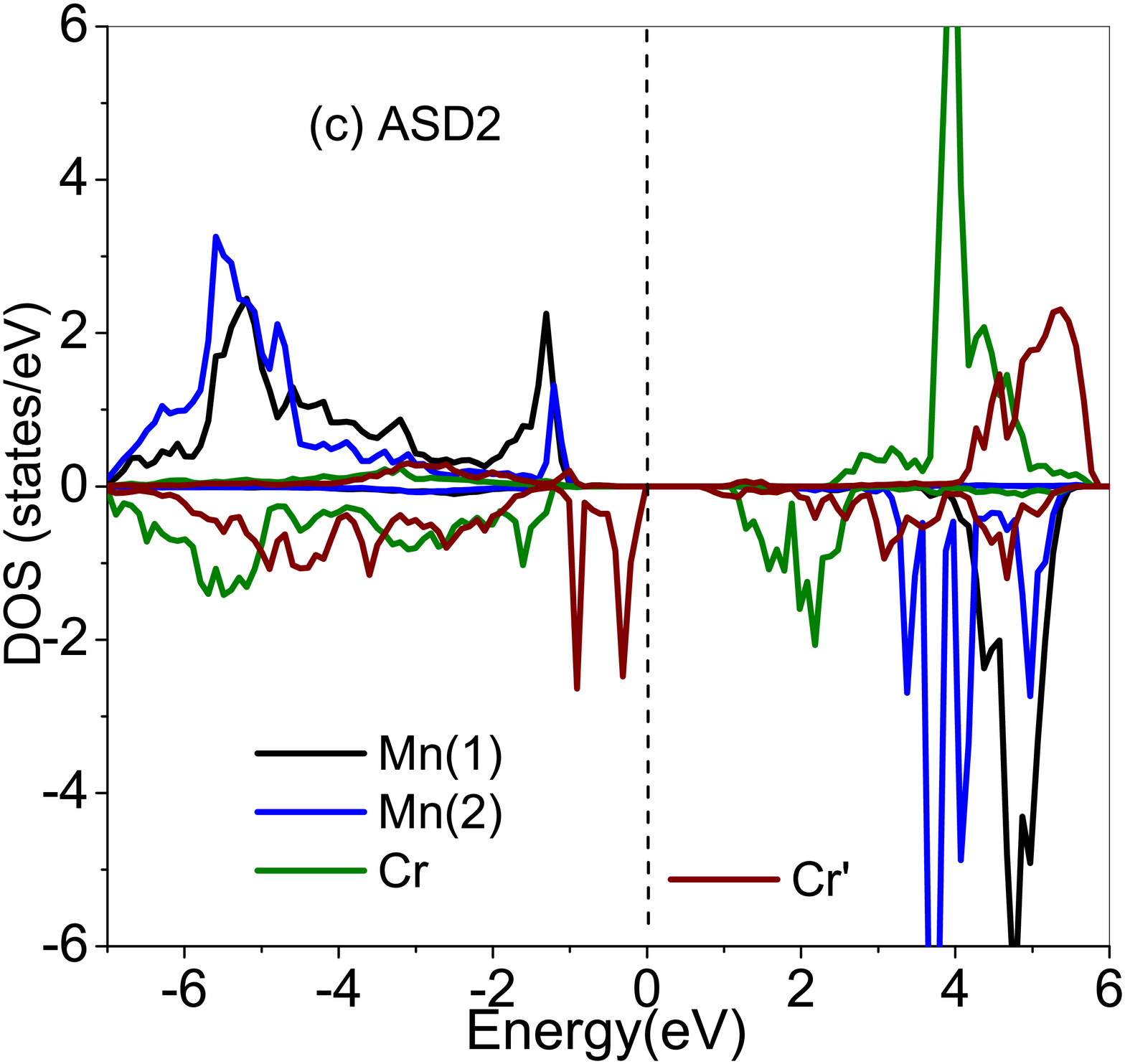}
\endminipage\hfill
\caption{FiM DOS of CaMnCrSbO$_{6}$ for $U_{eff}$$=$5 eV, in (a) ordered, (b) ASD1 and (c) ASD2 structures.}
\label{DOS_ferri_asd_Cr}
\end{figure}
\newline
We show the total $d$-orbital DOS of FiM CaMnCrSbO$_{6}$ with $U_{eff}$$=$5 eV in the regular structure (Fig.~\ref{DOS_ferri_asd_Cr}a) and ASD structures (Figs.~\ref{DOS_ferri_asd_Cr}b and \ref{DOS_ferri_asd_Cr}c).
The DOS of the regular structure is similar to that obtained for $U_{eff}$$=$3 eV, except for a larger band gap now, i.e. ${\sim}$2.1 eV.
The inclusion of ASD drastically affects the DOS.
In ASD1, the spin-up and spin-down DOS of the unswitched Mn and Cr atoms, are about 1 eV below $E_{F}$, as seen from Fig.~\ref{DOS_ferri_asd_Cr}b.
In comparision to the regular Cr and Mn atoms, the switched atoms ie Mn' and Cr', show drastically different spectral features in the DOS.
The spin-up DOS just below $E_{F}$ is dominated by the $d$ DOS of the Cr' atom, which shows sharp peaks at ${\sim}$-0.2 and ${\sim}$-0.5 eV.
 Similarly, the  spin-down DOS just below and above $E_{F}$ comprises of sharp peaks due to $d$ DOS of the Mn' atom, thereby making the band-gap almost negliblible in ASD1.
The ASD2 does not affect the magnetic moments, even though interestingly it partially affects the DOS and reduces the band gap.
As seen in Fig.~\ref{DOS_ferri_asd_Cr}c, in ASD2 structure, the band gap has reduced to ${\sim}$ 1 eV, due to the switched Cr' $d$-states, which contribute to the spin-down DOS just below $E_{F}$.
\newline
The total $d$-electron occupation number obtained for Mn(1), Mn(2) and Cr in the ordered supercell are 5.06, 5.03 and 4.23 respectively. 
While in Mn(1) and Mn(2), the $d$-occupation number is quite in agreement with the formal valency of +2 (3$d^{5}$), the $d$-occupancy of Cr is considerably higher than that expected from a valency of +3 (3$d^{3}$), thus indicating a greater charge transfer nature.  
In ASD1, the occupancies of the switched atoms, Mn' and Cr', show significant changes. The occupancy of Mn' has decreased to 4.88, while that of Cr' has increased to 4.40. 
The changes in occupation is thus reflected in the magnetic moment of the switched atoms. 
As seen from Table 3, the magnitude of Mn' magnetic moment has reduced to 3.96 ${\mu}_{B}$, while Cr' shows an enhanced magnetic moment of magnitude 3.96 ${\mu}_{B}$. These absolute values of magnetic moments indicate that the Mn' and Cr' atoms possess formal valencies of +3 and +2 respectively.
In ASD2, though the occupancy of Cr' increases to 4.34, a similar increase in Cr magnetic moment is not observed. As mentioned in Table 3, the magnitude of Cr' moment has slightly reduced to 3.06 ${\mu}_{B}$.
The inclusion of ASDs result in an overall decrease in the magnetic moment of Mn within the supercell, in accordance with the experiments. 
\subsection{Determination of exchange constants}
\begin{figure*}[!htb]
\centering
\includegraphics[scale=0.55]{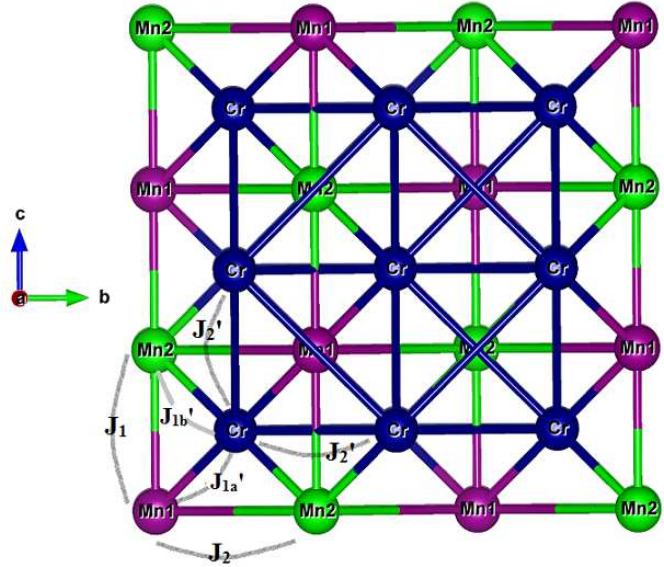}
\caption{Cr/Mn lattice with paths marked for various exchange interactions.}
\label{exchange_lattice}
\end{figure*}
The stability of the FiM ground state can be understood in a better manner with the help of the various exchange constants in the system. 
Since CaMnCrSbO$_{6}$ is found to be insulating from our electronic structure calculations, with a strong Mott Hubbard character, the exchange interactions between the two TM ions arise predominantly through the intervening O$^{2-}$ $p$-orbitals, thus resulting in superexchange interactions.
The exchange interactions ($J$'s) were evaluated by mapping the total energy of the various magnetic configurations (Table 2) to the Ising model Hamiltonian,
\begin{equation}
H  = -\sum_{i,j}J_{ij}S_{i}^{z}S_{j}^{z} 
\end{equation}
where $J_{ij}$ is the exchange interaction strength, while $S_{i}^{z}$ and $S_{j}^{z}$ correspond to Mn$^{2+}$ or Cr$^{3+}$ spins, as per applicability.
The $J$'s were obtained for the following Hubbard parameters, $U$$=$5.5 eV, $J_H$$=$1 eV for Cr and $U$$=$6 eV, $J_{H}$$=$1.5 eV for Mn. The role of oxygen and other non-magnetic atoms in the lattice has been neglected here. 
Due to presence of two different transition metals, there occur multiple exchange paths, as compared to the case of single and double perovskites.
The various exchange mechanisms can be understood from the magnetic lattice, whose projection onto the $b$$-$$c$ plane is shown in Fig.~\ref{exchange_lattice}.
In Table 4, we list the various $J$'s in order of increasing distance `$d$' for different sets of TMs in the magnetic lattice.  
\begin{table}
\caption{\label{tabone}\label{tabone}Exchange constants (meV) between the magnetic ions in CaMnCrSbO$_{6}$. The '+' and '-' sign indicate the FM or AFM nature of the exchange interaction.} 
\begin{center}
\lineup
\item[]\begin{tabular}{@{}*{8}{l}}
\br   
\0\0 & & Path&$d(\AA$) & Method I & Method II & Method III \cr
\0\0 & &&&\m(meV)&\m(meV)&\m(meV)\cr
\mr
\0\0  & $J_{1a}'$  &Mn(1)-Cr   &3.3031 &\m -0.41 &\m\m - &\m -0.46   \cr
\0\0  & $J_{1b}'$  &Mn(2)-Cr   &3.3031 &\m -0.06 &\m\m - &\m -0.14   \cr
 \0\0& $J_{1}$   &Mn(1)-Mn(2)($c$-axis)   &3.8104 &\m -0.07 &\m -0.11 &\m -0.11 \cr
\0\ & $J_{2}$  &  Mn(1)-Mn(2)($a$$-$$b$) &5.3966 &\m -0.03 &\m -0.05 &\m -0.05  \cr
\0\0&$J_{2}'$ &Cr-Cr&5.3966&\m +0.32&\m +0.37 &\m +0.28 \cr
\br
\end{tabular}
\end{center}
\end{table}
\newline
Considering only the magnetic atoms, we note that Cr atoms are the nearest neighbours to both, Mn(1) and Mn(2) atoms. Each Cr atom has two Mn(1) and two Mn(2) atoms as the nearest neighbours. Due to different ligand environment of the Mn(1) and Mn(2) atoms, we expect the Mn(1)-Cr and Mn(2)-Cr exchange interactions as distinct. 
The Mn(1) atoms are flanked by two Mn(2) atoms along the $c$-axis as nearest neighbours and vice versa. In the $a$$-$$b$ plane, each Mn(1) is surrounded by four Mn(2) atoms and vice-versa.
The Mn(1)-Mn(2) separation in the $a$$-$$b$ plane is much larger and is the same as the Cr-Cr separation.
For each Cr atom, there are twelve next nearest neighbour Cr atoms, with four Cr atoms in the (110), (101) and (011) planes, each. The corresponding Cr-Cr interactions are denoted as $J_2^{110'}$, $J_2^{101'}$ and $J_2^{011'}$, respectively.
Substituting $S_{Mn}^{z}$$=$${\pm}$5/2 and $S_{Cr}^{z}$$=$${\pm}$3/2 for the Mn$^{2+}$ and Cr$^{3+}$ spins, respectively, the Ising Hamiltonian is expanded for the different magnetic configurations.
After a careful enumeration, the total energy in the FM and FiM configurations can be expanded as,
\begin{equation}
E^{FM}   = -25J_{1}-50J_{2}-18J_2^{110'}-18J_2^{101'}-18J_2^{011'}-30J_{1a}'-30J_{1b}'
\end{equation}
\begin{equation}
E^{FiM}   = -25J_{1}-50J_{2}-18J_2^{110'}-18J_2^{101'}-18J_2^{011'}+30J_{1a}'+30J_{1b}'
\end{equation}
To obtain Mn(1)-Cr and Mn(2)-Cr exchange constants, i.e. $J_{1a}'$ and $J_{1b}'$, unambiguously, two additional calculations were performed. In both the calculations, among the four Cr$^{3+}$ spins in the unit cell, a single Cr$^{3+}$ spin was fixed as spin-down. In the first configuration (denoted as configuration $M$), a single Mn(1)$^{2+}$ spin was kept as spin-down, and in the second case (denoted as configuration $N$), a single Mn(2)$^{2+}$ spin was fixed as spin-down. Due to cancellation of most of the terms, the total energy for configurations $M$ and $N$ reduces to,
\begin{equation}
\begin{split}
E_{M} & = -15J_{1b}' \\
E_{N} & = -15J_{1a}'
\end{split}
\end{equation}
Our calculations show that the energies of both the configurations $M$ and $N$ are higher than that of FM by +8.33 and +4.93 meV per unit cell, respectively.
From equations (2), (3) and (4), we obtain $J_{1a}'$$=$-0.41 meV and $J_{1b}'$$=$-0.06 meV.
The negative values indicate that these exchange interations are anti-ferromagnetic in nature. 
Interestingly, the differences in the local environment of both the Mn$^{2+}$ ions result in a large difference between Mn(1)-Cr and Mn(2)-Cr interaction strengths. To obtain the exchange interactions within the individual Mn and Cr sublattices, we refer to the calculations in which a) Mn sublattice was kept fixed in FM configuration and magnetic configuration of the Cr sublattice was varied, and b) Cr sublattice was kept fixed in FM configuration, while the magnetic configuration of the Mn sublattice was varied (Table 2).
The set of equations for the A-, C- and G-type arrangement of the Mn-sublattice is,
\begin{equation}
E_{Mn}^{A}   = 25J_{1}-50J_{2}-18J_2^{110'}-18J_2^{101'}-18J_2^{011'}
\end{equation}
\begin{equation}
E_{Mn}^{C}   = -25J_{1}+50J_{2}-18J_2^{110'}-18J_2^{101'}-18J_2^{011'}
\end{equation}
\begin{equation}
E_{Mn}^{G}   = 25J_{1}+50J_{2}-18J_2^{110'}-18J_2^{101'}-18J_{2}^{011'}
\end{equation}
Similarly, the set of equations for A-, C- and G-type arrangement of the Cr-sublattice is,
\begin{equation}
E_{Cr}^{A}   = -25J_{1}-50J_{2}-18J_2^{110'}+18J_2^{101'}+18J_2^{011'}
\end{equation}
\begin{equation}
E_{Cr}^{C}   = -25J_{1}-50J_{2}+18J_2^{110'}-18J_{2}^{101'}+18J_2^{011'}
\end{equation}
\begin{equation}
E_{Cr}^{G}   = -25J_{1}-50J_{2}+18J_2^{110'}+18J_2^{101'}-18J_2^{011'}
\end{equation}
After solving the equations (2)-(10), it is found that all the Mn(1)-Mn(2) interactions are anti-ferromagnetic and smaller as compared to the Cr-Cr interactions which are found to be ferromagnetic. While, $J_2^{110'}$$=$+0.37 meV, both $J_2^{101'}$, and $J_2^{011'}$ are found to be +0.29 meV, suggesting a small anisotropy. However, we only consider the average of the three values, denoted as $J_2'$. Inspite of having nearly the same separation distance, the Cr-Cr interaction strength in our system is much greater than Cr-Cr exchange strength of +0.05 eV, obtained in Ca$_{2}$CrSbO$_{6}$\cite{Baidya_PRB2012}. 
In Table (4) we list all these exchange constants in the column under Method I. 
\newline
To assertain the values of $J$'s obtained above, we have independently obtained the exchange parameters, using two additional methods, labelled as Method II and III.  In Method II, one magnetic element is entirely replaced by a non-magnetic element. Hence, to obtain the Cr-Cr and Mn-Mn exchange interactions, the Mn(Cr) atoms are replaced by isovalent Zn(Sc) atoms. 
However, in this method, the inter sublattice i.e. Mn-Cr exchange interaction strength cannot be determined. The intra-sublattice exchange interactions thus obtained are shown in Table 4 in the column Method II.
In Method III, except for the two magnetic atoms whose exchange interaction strength is to be determined, rest of the magnetic atoms are replaced by the isovalent non-magnetic equivalents\cite{Weingart_PRB2012}.
The corresponding exchange interaction strength is then proportional to the energy difference between the FM and AFM arrangement of the two spins and are listed in Table 4 under Method III. 
The sign of $J$ for the various exchange interactions, obtained from the three methods are the same, and the magnitudes almost equal, though discrepencies upto 90 ${\mu}$eV are found in the case of Cr-Cr interactions.
It is noted that these values of nearest neighbour $J$ from DFT studies for this double-double perovskite are an order of magnitude smaller than those found in single and double perovskite compounds for similar values of $U$ and $J_{H}$\cite{Weingart_PRB2012, Franchini_PRB2016, Gauvin-Ndiaye_PRB2018}. 
The Mn(1)-Mn(2) exchange interactions which are found to be anti-ferromagnetic, appears contrary to the experimental studies. 
The ferrimagnetic behaviour obtained from bulk magnetization and neutron diffraction suggests a parallel alignment of Mn$^{2+}$ spins which is possible only when the Mn(1)-Mn(2) exchange interaction is ferromagetic.
To understand the origin and sign of the various exchange interactions, we have investigated the superexchange mechanisms in the preceeding section.
\begin{figure}[!htb]
\minipage{0.55\textwidth}
 \includegraphics[width=\linewidth]{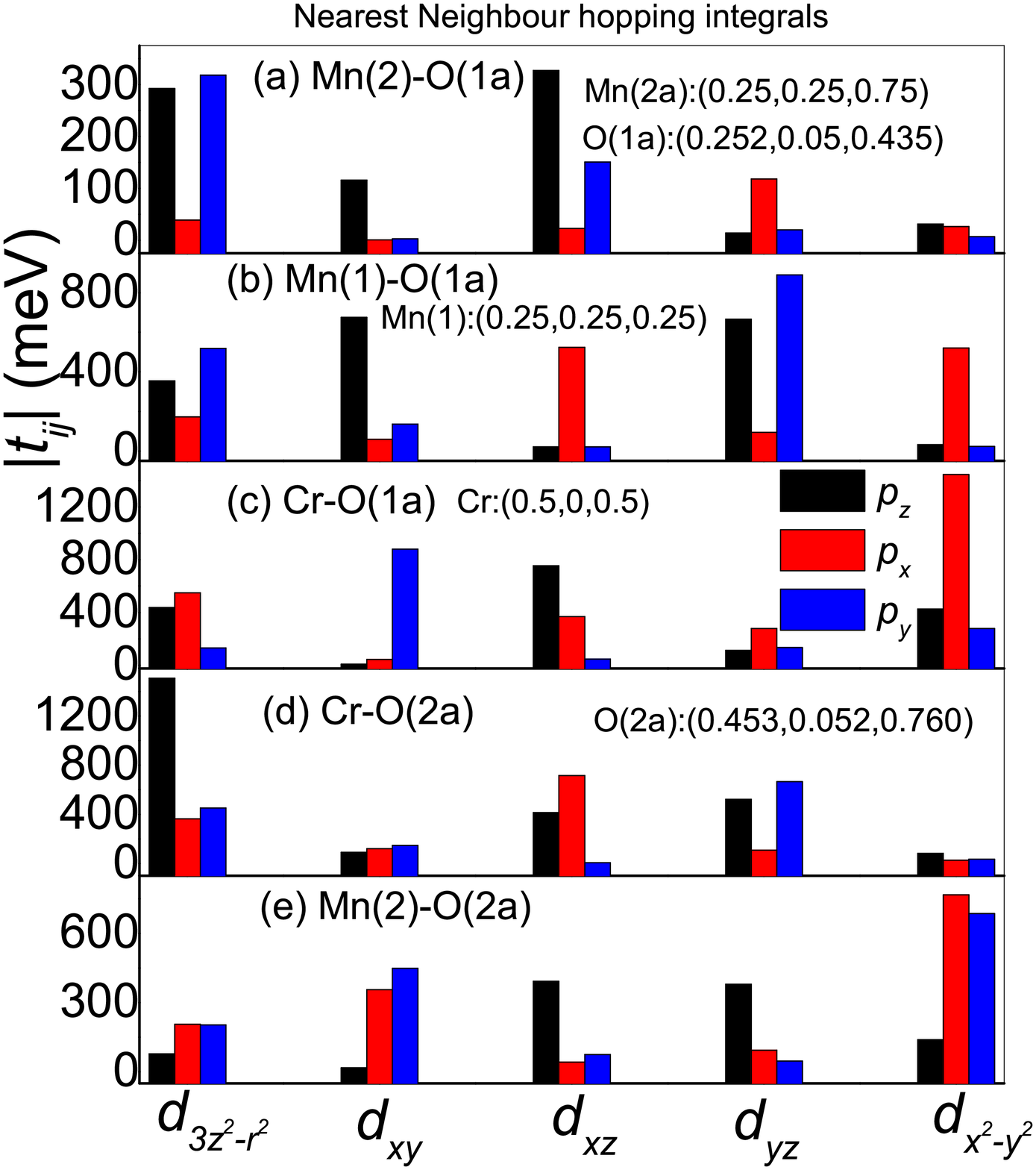}
\endminipage\hfill
\minipage{0.55\textwidth}
 \includegraphics[width=\linewidth]{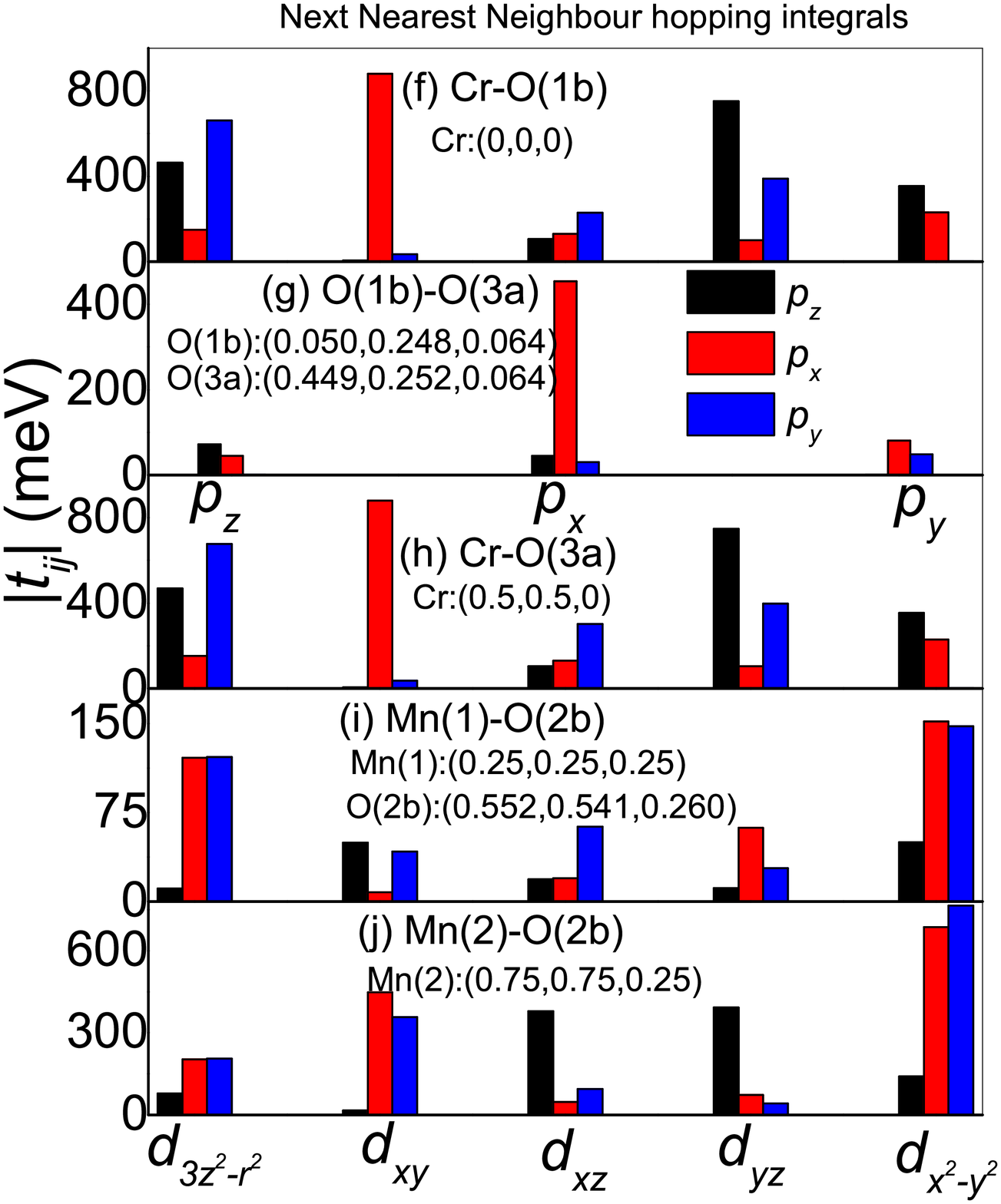}
\endminipage\hfill
\caption{Magnitude of hopping integrals between the MLWFs of $d$ orbitals of Mn(1), Mn(2) and Cr with the O $p$ orbitals. Hopping integrals correspond to following superexchanges, (a)-(b) Mn(1)-O-Mn(2) (along the $c$-direction), (b)-(c) Mn(1)-O-Cr, (d)-(e) Mn(2)-O-Cr, (f)-(h) Cr-O-O-Cr `super-superexchange' (i)-(j) Mn(1)-O-Mn(2) ($a$$-$$b$ plane). The coordinates of the transition metal and Oxygen atoms are mentioned for each panel.}
\label{Hopping_integrals}
\end{figure}
\subsection{Superexchange from hopping integrals}
\begin{figure}
\minipage{0.30\textwidth}
 \includegraphics[width=\linewidth]{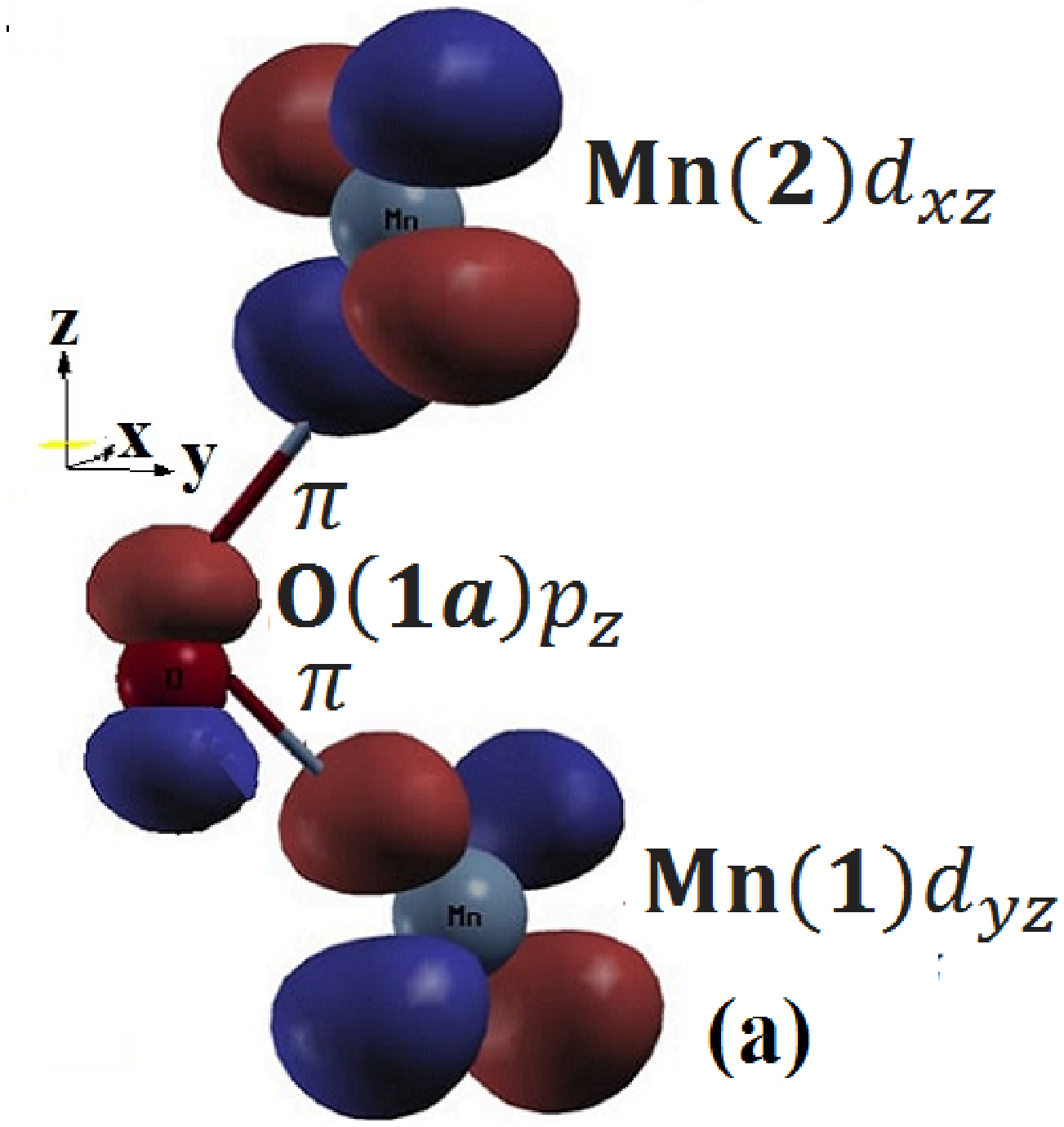}
\endminipage\hfill
\minipage{0.34\textwidth}
 \includegraphics[width=\linewidth]{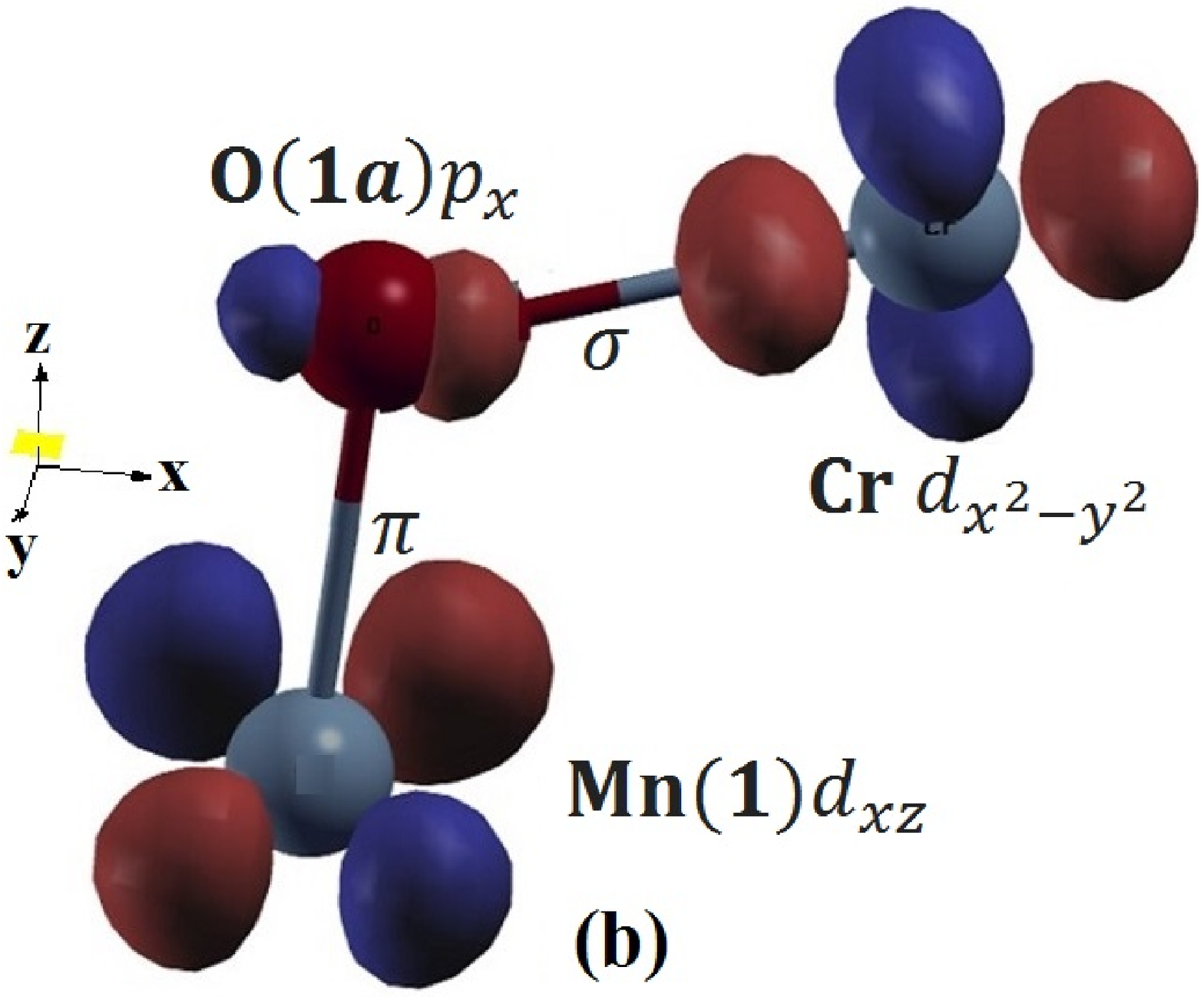}
\endminipage\hfill
\minipage{0.34\textwidth}
 \includegraphics[width=\linewidth]{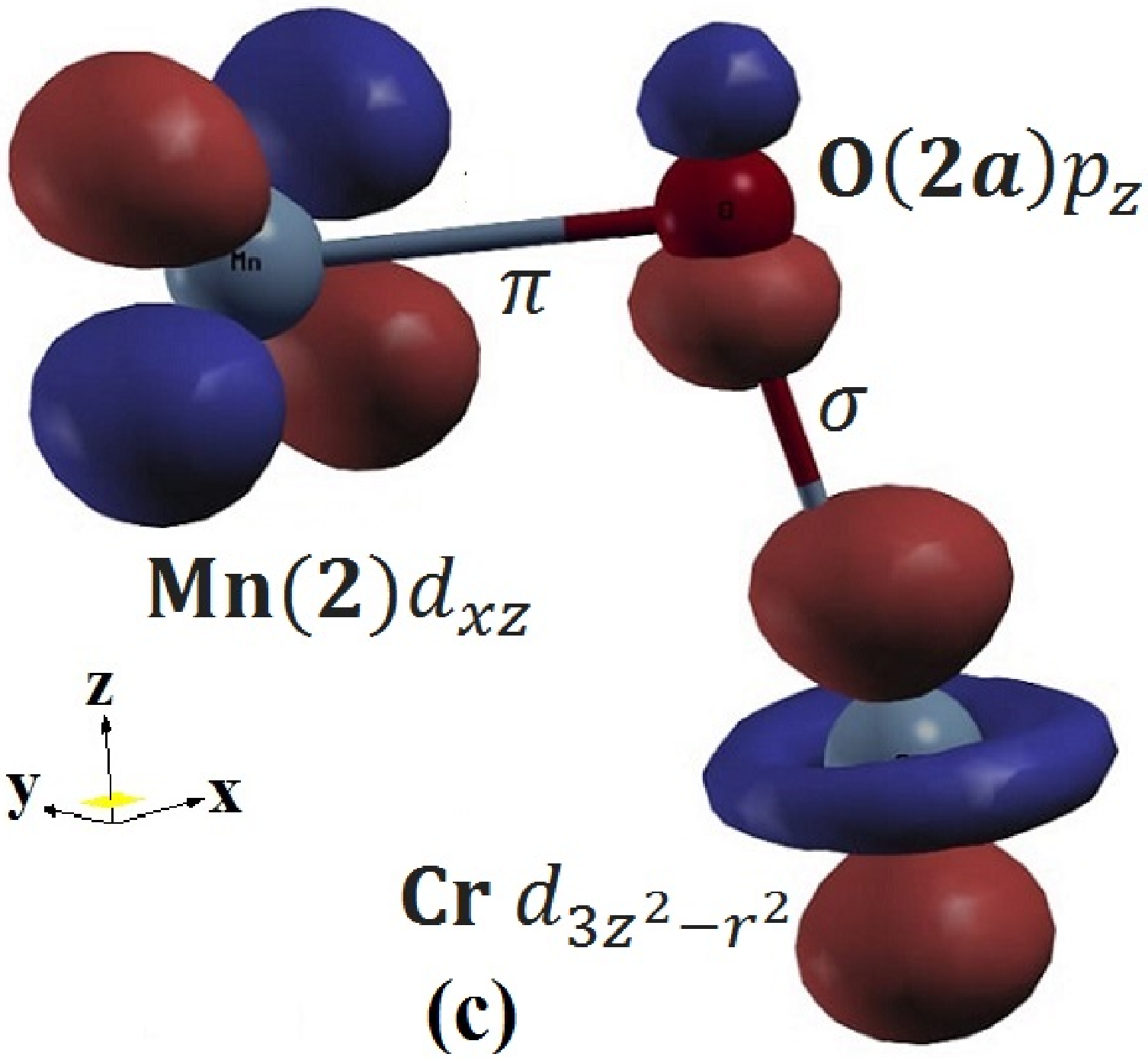}
\endminipage\hfill
\minipage{0.32\textwidth}
 \includegraphics[width=\linewidth]{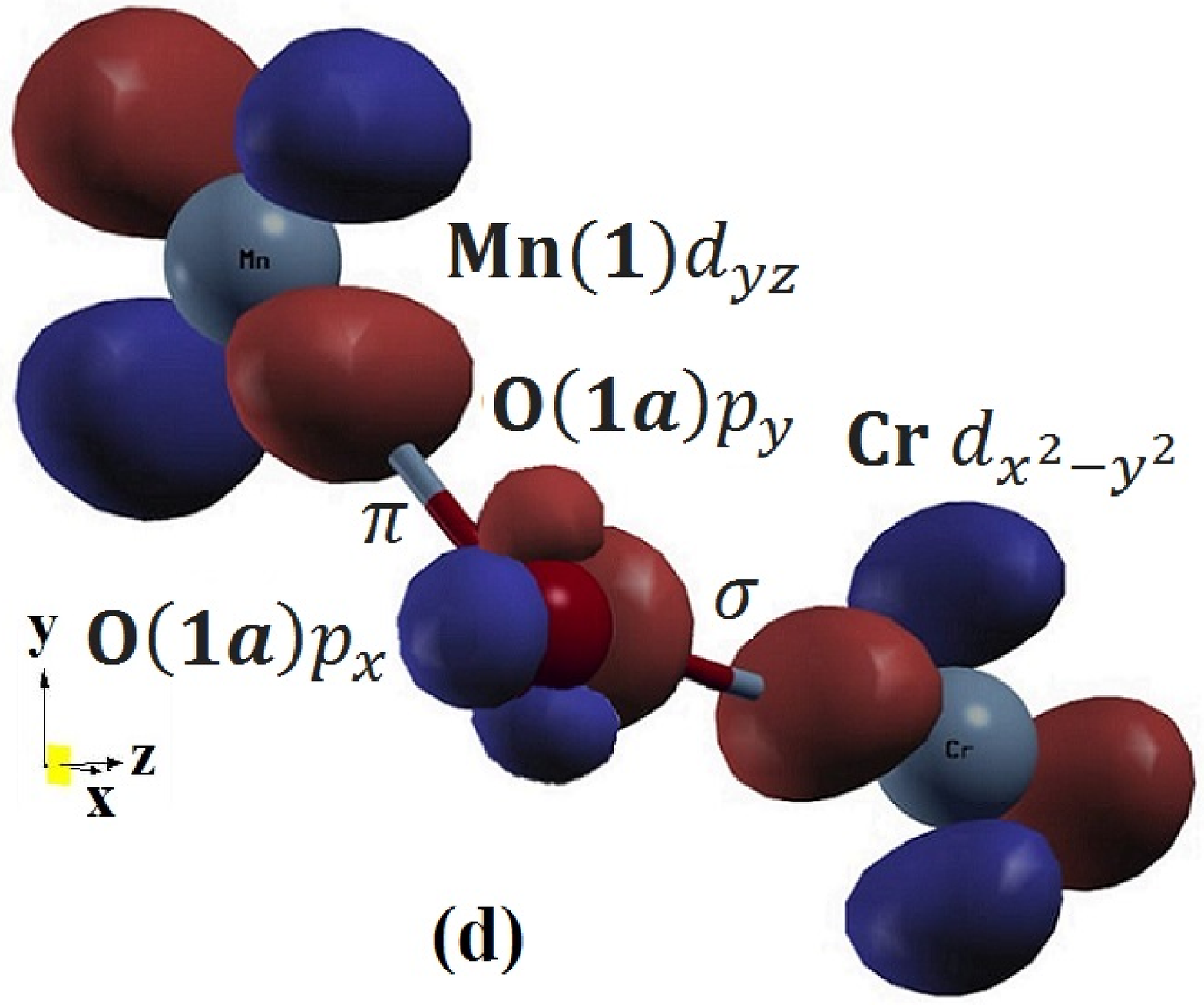}
\endminipage\hfill
\minipage{0.32\textwidth}
 \includegraphics[width=\linewidth]{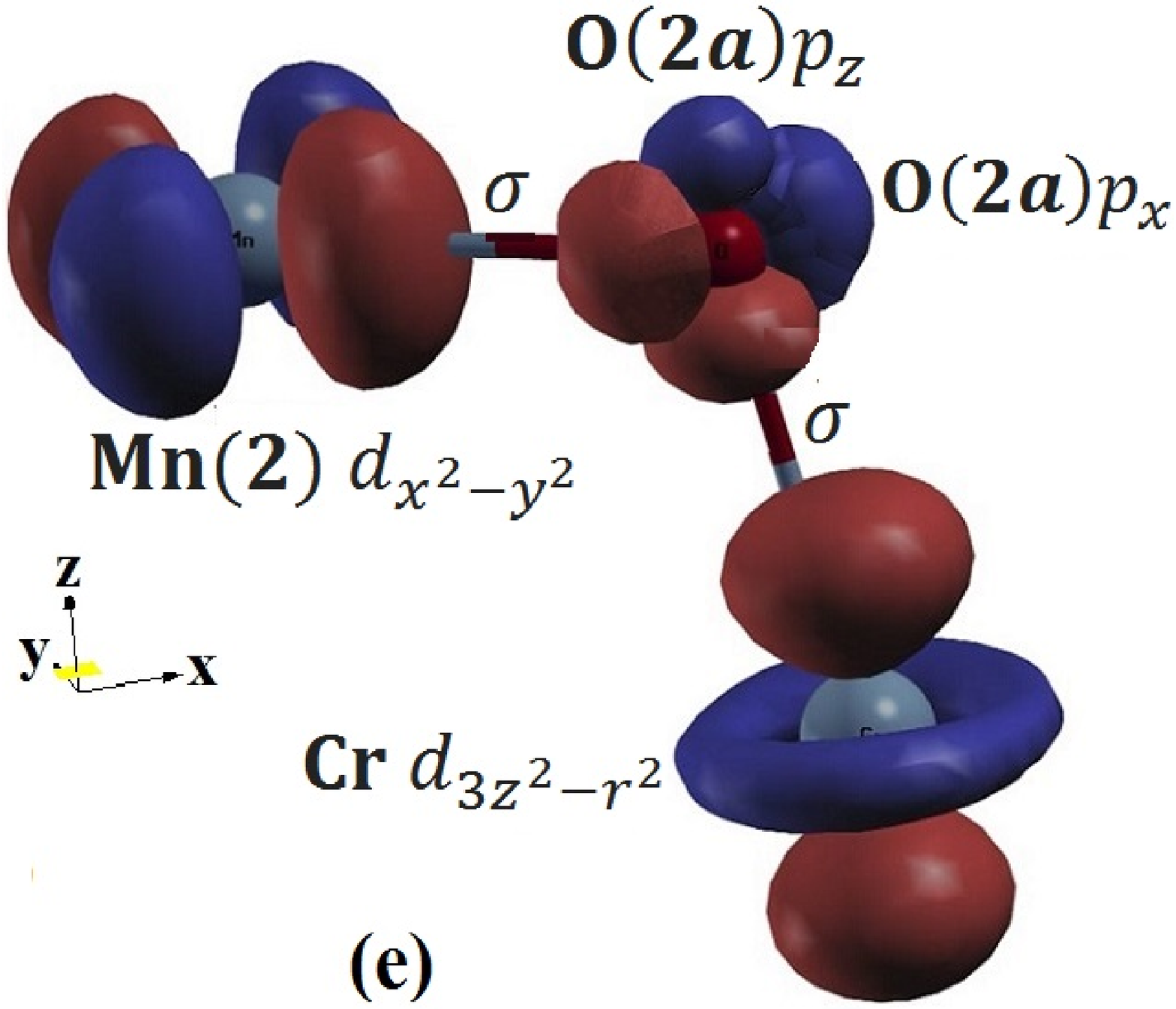}
\endminipage\hfill
\minipage{0.30\textwidth}
 \includegraphics[width=\linewidth]{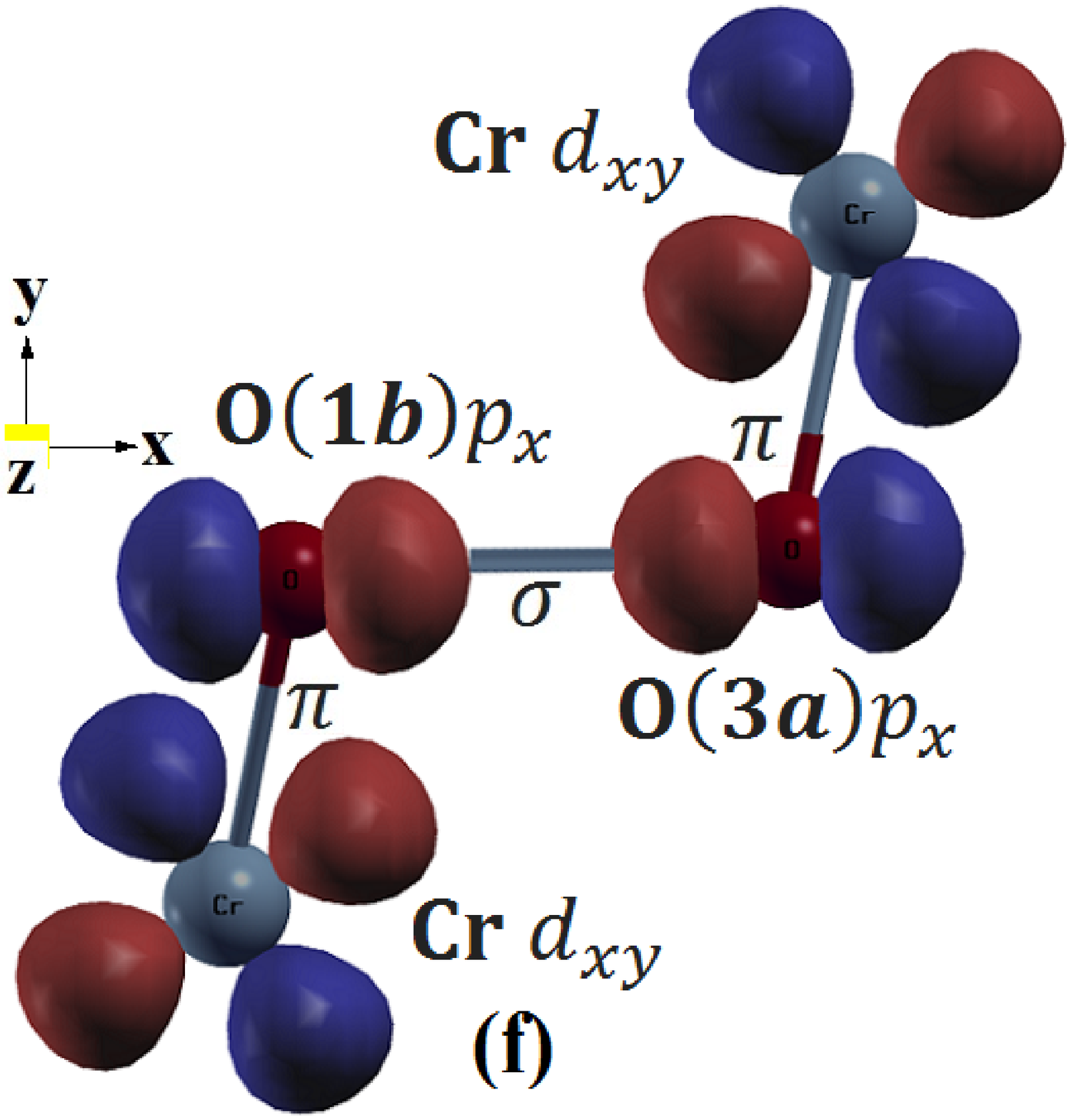}
\endminipage\hfill
\caption{Exchange paths between selected atomic Mn, Cr $d$ orbitals via O $p$ orbitals, obtained from Wannier function. The nature of the bonding (${\sigma}$ or ${\pi}$) is also highlighted. (a) Nearest neighbour (along $c$-direction) Mn(1)-O(1a)-Mn(2) AFM superexchange; (b) and (c) AFM superexchange from Mn(1)-O(1a)-Cr and Mn(2)-O(2a)-Cr. (d) and (e) FM superexchange involving two oxygen orbitals and (f) FM superexchange between the two Cr atoms involving two intermediate O atoms.}
\label{super-exchange_path}
\end{figure}

\begin{figure*}[!htb]
  \centering
 \includegraphics[scale=0.30]{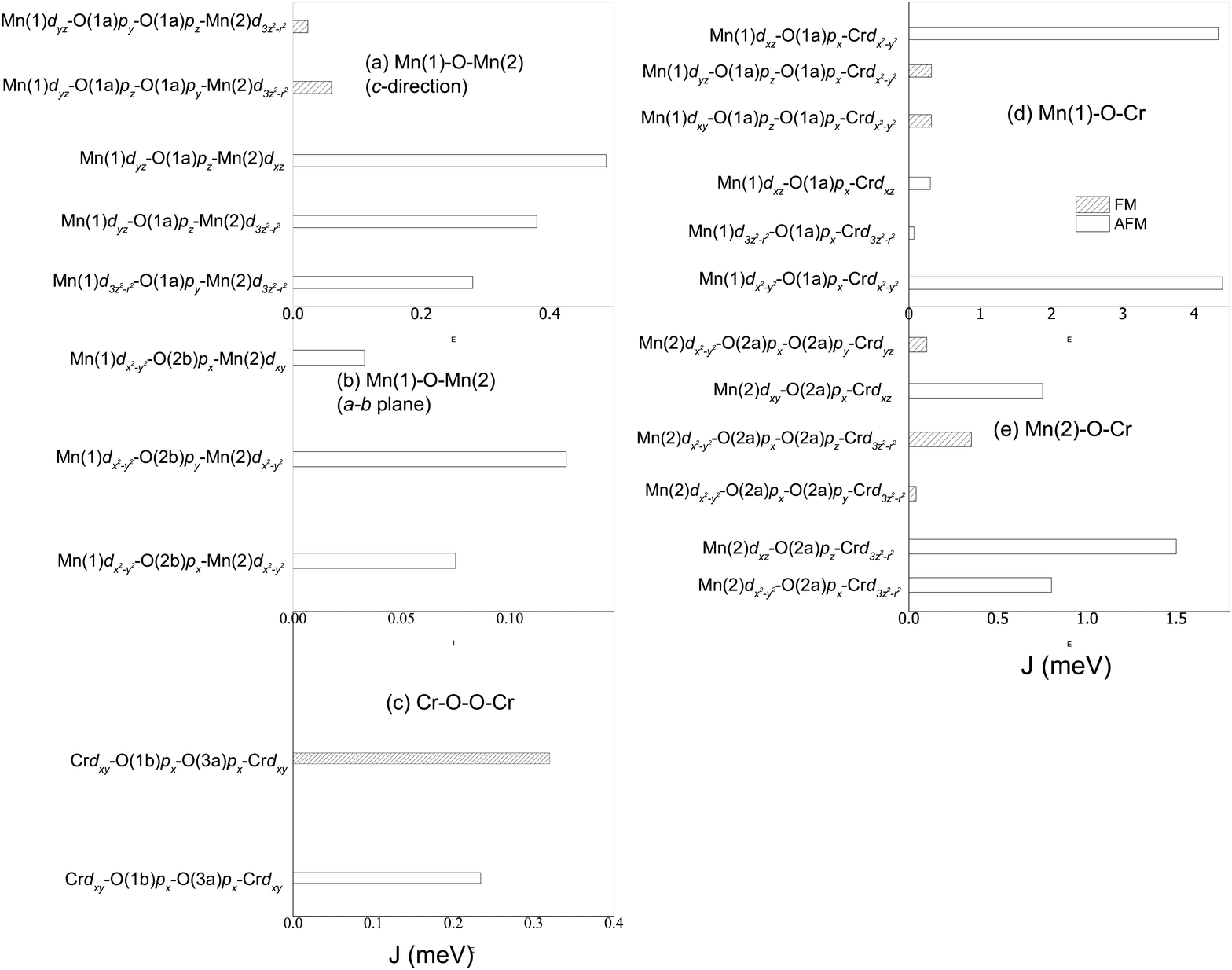}
\caption{ Magnitude of calculated exchange interaction energies ($x$-axis) for various kinds of superexchange interactions calculated from equations (11)-(13). The corresponding exchange paths are mentioned on the $y$-axis.}
\label{J_values}
\end{figure*}

The exchange interaction strengths obtained in the preceding section can be understood in terms of the hopping amplitudes, between the Wannier orbitals of Mn(1), Mn(2) and Cr via the O $p$-orbitals. The $J$'s have been calculated from the overlap integrals between the metal and the ligand Wannier functions.
As shown in Fig.~\ref{dosnm_cef}(d-f) and Fig.~\ref{Wannier}, due to the differences of the point group symmetry and the local coordinate axes of Mn(1)$^{2+}$, Mn(2)$^{2+}$ and Cr$^{3+}$ ions, the nearest neighbour $d$-$p$ hoppings show ${\sigma}$ as well as ${\pi}$ character.
\newline
In Fig.~\ref{Hopping_integrals}(a-e), we show the magnitudes of the $d$$-$$p$ hopping integrals ($t_{pd}$), which give rise to the superexchange interaction between the nearest neighbour Mn(1)-Mn(2), Mn(1)-Cr and Mn(2)-Cr spin pairs, while Fig.~\ref{Hopping_integrals}(i) and \ref{Hopping_integrals}(j) correspond to $d$$-$$p$ hopping associated with the in-plane Mn(1)-Mn(2) superexchange interactions.
Fig.~\ref{Hopping_integrals}(f-h) corresponds to the Cr-Cr exchange occuring via two intermediate O$^{2-}$ ions, and is labelled as 'super-superexchange'. Thus, while Figs.~\ref{Hopping_integrals}(f) and \ref{Hopping_integrals}(h) correspond to the hopping between the adjacant Cr-O pairs, Fig.~\ref{Hopping_integrals}(g) represents the hopping between the two O$^{2-}$ ions, denoted by $t_{pp}$.
The coordinates of the TM and the intermediate oxygen atoms are also mentioned in the sub-plots. The labels (1a), (2a),.. have been placed only for the purpose of identification of the various O atoms. 
Based on these amplitudes of the $d$$-$$p$ hoppings, we illustrate some of the possible paths for the nearest neighbour Mn(1)-O(1a)-Mn(2), Mn(1)-O(1a)-Cr, Mn(2)-O(2a)-Cr superexchange interactions in Fig.~\ref{super-exchange_path}(a-e).
The Mn(1)-O-Mn(2) has a bond-angle of ${\sim}$100$^{o}$, while the Mn(1)-O-Cr and Mn(2)-O-Cr bond angles are ${\sim}$105$^{o}$; these can thus be understood in the formalism of the 90$^{o}$ superexchange according to the Goodenough-Kanamori rules\cite{Kanamori_1959}. 
With the ${\sigma}$ as well as ${\pi}$ hopping contributing to the superexchange process, the 90$^{o}$ superexchange is more complex, though
weaker than the 180$^{o}$ antiferromagnetic superexchange which is seen in the perovskite systems\cite{Kanamori_1959}.
\newline
In general, when the hopping between the $d$-orbitals of two TM ions occurs through a single doubly occupied O$^{2-}$ $p$-orbital, the resultant superexchange is antiferromagnetic. Within a simple formalism, considering only single $d$-orbital at the TM site the expression for $J$, derived from a three site Hubbard model (Appendix, equations (A1)-(A6)) is, 
\begin{equation}
J_{AFM}=-2(t_{pd1}t_{pd2})^{2}\left[\frac{1}{{\tilde U}_{1}^2(U_{1}+{\Delta}{\epsilon})}+\frac{1}{{\tilde U}_{2}^2(U_{2}-{\Delta}{\epsilon})}+ \frac{1}{{\bar U}}\left(\frac{1}{{\tilde U}_{1}}+\frac{1}{{\tilde U}_{2}}\right)^{2}\right]
\end{equation}
Here, the different hopping strengths, $U$ and charge trasfer energies(${\Delta}_{pd}$) associated with two different TM ions are taken into account. The values of Hubbard $U$ for Mn and Cr are fixed at 6 and 5.5 eV, respectively. The various terms and symbols used are explained in the Appendix. Though valid for metal-ligand-metal in 180$^{o}$ geometry, the hopping parameters take into account the deviation from the ideal geometry. We now explain the various superexchange processes.
%
%
%
%
%
\newline
The superexchange between Mn(1)$d_{yz}$ and Mn(2)$d_{xz}$ via O(1a)$p_{z}$ is shown in Fig.~\ref{super-exchange_path}a. 
Since, the Mn(2)-O(1a) bondlength (${\sim}$2.83 {\AA}) is greater than the Mn(1)-O(1a) bondlength (${\sim}$2.03 {\AA}), the Mn(2)$d$-O(1a)$p$ hopping amplitude is smaller, thus resulting in a small Mn(1)-O(1a)-Mn(2) exchange strength. 
Using equation (11), we obtain the anti-ferromagnetic $J_{AFM}$ value of -0.5 meV for this particular superexchange path, which is the maximum value for the Mn(1)-O-Mn(2) superexchange, along the $c$-direction. A comparable $J_{AFM}$ value of -0.4 meV is obtained for the Mn(1)$d_{yz}$-O(1a)$p_{z}$-Mn(2)$d_{3z^2-r^2}$ superexchange.
 The antiferromagnetic $J_{AFM}$ calculated for a few combinations of Mn(1), Mn(2) $d$ and O(1a) $p$-orbitals are listed in Fig.~\ref{J_values}a. 
The superexchange interaction between Mn(1) and Mn(2) atoms, situated in-plane, occurs through the intermediate O(2b) atom.
As seen from Figs.(\ref{Hopping_integrals}i and \ref{Hopping_integrals}j), the Mn(1)-O(2b) hopping amplitude is smaller than that for Mn(2)-O(2b).
The $J_{AFM}$ values for various possible Mn(1)-O-Mn(2) in-plane superexchange paths are shown in Fig.~\ref{J_values}b.
The highest value of $J_{AFM}$$=$-0.12 meV is obtained for the Mn(1)$d_{x^2-y^2}$-O(2b)$p_{x}$-Mn(2)$d_{x^2-y^2}$ superexchange which is close to the value obtained from DFT.
\newline
The Mn(1)-Cr exchange can be understood through the superexchange via O(1a) $p$-orbitals. 
Due to ${\sigma}$-bond, between Cr$d_{x^2-y^2}$ and O(1a)$p_{x}$, the  hopping amplitude is the maximum (Fig.~\ref{Hopping_integrals}c).
Out of the many possible superexchange mechanisms involving the different $d$-orbitals of Mn(1) and Cr and the $p$-orbitals of O(1a), we show the path Mn(1)$d_{xz}$-O(1a)$p_{x}$-Cr$d_{x^2-y^2}$ (Fig.~\ref{super-exchange_path}b), which yields a large antiferromagnetic superexchange strength ($J_{AFM}$$=$-4.2 meV). Similar value of $J_{AFM}$ is obtained, for instance, in the case of Mn(1)$d_{x^2-y^2}$-O(1a)$p_{x}$-Cr$d_{x^2-y^2}$ superexchange (Fig.~\ref{J_values}d). Thus the maximum $J_{AFM}$ values (Fig.~\ref{J_values}d) are an order of magnitude higher than the exchange strength ($J_{1a}'$) directly obtained from DFT (Table 4).
\newline
The Mn(2)-Cr exchange can be understood through the superexchange via the $p$-orbitals of O(2a).
The Mn(2)-O(2a) hopping amplitudes are smaller, as compared to those for Cr-O(2a), as shown in Figs.~\ref{Hopping_integrals}(e) and ~\ref{Hopping_integrals}(d), respectively. 
In Fig.~\ref{super-exchange_path}c we show the path Mn(2)$d_{xz}$-O(1a)$p_{x}$-Cr$d_{3z^2-r^2}$, which yields an antiferromagnetic superexchange strength ($J_{AFM}$$=$-1.5 meV).
In Fig.~\ref{J_values}e, we show $J_{AFM}$ for some of the AFM superexchange interactions for Mn(2)-O-Cr.
For all these possible cases considered, the maximum $J_{AFM}$ for Mn(2)-O(2a)-Cr is nearly one-third of the maximum $J_{AFM}$ obtained for Mn(1)-O(1a)-Cr (Fig.~\ref{J_values}d), thus in agreement with our DFT calculations.
\newline
In a 90$^{o}$ superexchange as is present in our system, ferromagnetic superexchange between the TM ions also occurs due to hopping via two different $p$-orbitals of the single O$^{2-}$ ion\cite{Kanamori_1959}. Considering the additional $p$-orbital and an exchange coupling between the two O$p$-orbitals, the exchange interaction strength (Appendix B) is,
\begin{equation}
J_{FM} = \frac{\left(\frac{t_{pd1}^{2}}{{\tilde U}_{1}}+\frac{t_{pd2}^{2}}{{\tilde U}_{2}}\right)^{2}}{{\tilde U}-J_{p}}+\frac{\left({t_{pd1}^{2}}+{t_{pd2}^{2}}\right)^{2}}{J_{p}^{2}-{\tilde U}^{2}} 
\begin{bmatrix}
\frac{{\tilde U}}{{\tilde U}_{1}^2} & \frac{-J_{p}}{{\tilde U}_{1}{\tilde U}_{2}} \\
\frac{-J_{p}}{{\tilde U}_{1}{\tilde U}_{2}} & \frac{{\tilde U}}{{\tilde U}_{2}^{2}} \\
\end{bmatrix}
\end{equation} 
%
Using the above formula, the ferromagnetic superexchange strength for Mn(1)$d_{yz}$ and Cr$d_{x^2-y^2}$ occurring via O(1a) $p_{x}$ and $p_{y}$-orbitals (Fig.~\ref{super-exchange_path}d) yields a value, $J_{FM}$$=$ 0.5 meV (Fig.~\ref{J_values}d) which is nearly an order of magnitude smaller than the largest $J_{AFM}$. 
The Mn(2)$d_{x^2-y^2}$-O(2a)$p_{x}$-O(2a)$p_{z}$-Cr$d_{3z^2-r^2}$ superexchange (Fig.~\ref{super-exchange_path}e), which involves hopping through two ${\sigma}$ orbitals, yields $J_{FM}$$=$0.5 meV which however is comparable to $J_{AFM}$ of Mn(2)-O(2a)-Cr (Fig.~\ref{J_values}e). 
Similarly we obtain $J_{FM}$$=$0.1 meV for the Mn(1)$d_{yz}$-O(1a)$p_{z}$-O(1a)$p_{y}$-Mn(2)$d_{3z^2-r^2}$ ferromagnetic superexchange(Fig.~\ref{J_values}a).
Though ${J_{FM}}$ is always smaller than $J_{AFM}$, this can cause a net decrease in the overall strength of the antiferromagnetic interactions present in our system.
\newline
Finally, we discuss the `super-superexchange' between two Cr$^{3+}$ ions, which in principle is weaker than the rest of the interactions.
%
%
%
Due to the identical orthorhombic ligand symmetry, the Cr(0,0,0)-O(1b) hopping is almost identical to the Cr(0.5,0.5,0)-O(3a) hopping (Fig.~\ref{Hopping_integrals}f and h). As shown in Fig.~\ref{Hopping_integrals}g, maximum hopping occurs between the $p_{x}$ orbitals of O(1a) and O(3b), thus acting as a bridge for the `super-superexchange' process. 
%
%
Using sixth order perturbation theory, Toyoda {\it et al.}\cite{Toyoda_PRB2013,Toyoda_PRB2015} have calculated the exchange constant for the super-superexchange process ($J_{SS}$$=$-$t_{pd}^{4}$$t_{pp}^{2}$${\Delta}_{pd}^{-4}$(${\Delta}_{pd}^{-1}$+$U^{-1}$)).  
Considering only a single path Cr$d_{xy}$-O(1b)$p_{x}$-O(3a)$p_{x}$-Cr$d_{xy}$ (Fig.~\ref{super-exchange_path}f) which has maximum hopping amplitudes throughout, we obtain $J_{SS}$ $=$-0.25 meV.
So this formula predicts a correct order of magnitude, however, the sign of $J_{SS}$ is anti-ferromagnetic. This discrepency is due to the fact that the expression is obtained from the assumption that the ground state is made up of empty ligand orbital and a single hole at each TM ion site. This being appropriate for the cuprate superconductors (Cu$^{2+}$, $d^{9}$) or Cr$^{2+}$($d^{4}$) systems, but not suitable for our system\cite{Jefferson_PRB1993}.
Hence, similar to the case of first order superexchange, we extended the formalism to include the second O$^{2-}$ ion and obtained expression for the `super-superexchange' constant ${\tilde J_{SS}}$ (Appendix C) as,
\begin{equation}
{\tilde J_{SS}}=\frac{2t_{pd}^2(t_{pd}^2+t_{pp}^2)}{{\tilde U}({\tilde U}+t_{pd})^{2}}-\frac{2t_{pd}^{4}t_{pp}^{2}}{U{\tilde U}^{4}}-\frac{2t_{pd}^{4}}{{\tilde U}^{3}}+\frac{4t_{pd}^{4}t_{pp}^{2}}{{\tilde U}^{5}}
\end{equation}
Using equation(13), we obtain $J_{SS}$$=$0.075 meV for Cr$d_{xy}$-O(1b)$p_{x}$-O(3a)$p_{x}$-Cr$d_{xy}$ path, which though smaller has the right ferromagnetic character. However, on neglecting $t_{pd}$ in comparision to ${\tilde U}$ in the first term, we obtain a higher value of${\tilde J_{SS}}$$=$0.32 meV, which is close to the value obtained from DFT and hence can be considered as an upper limit.
The maximum ${\tilde J_{SS}}$ and $J_{SS}$ are shown in Fig.~\ref{J_values}c.
%
%
%

%
\subsection{Determination of Curie temperature from mean field theory}
Finally, based on the $J$-values obtained from our DFT calculations, we have determined the Curie temperature of CaMnCrSbO$_{6}$ using mean field theory.
Similar to Section 4.4, we describe the Mn/Cr sublattices as a system of Ising spins.
Due to the anti-ferromagnetic Mn(1)-Mn(2) interactions within the $a$$-$$b$ plane and $c$-directions, a two-sublattice system followed by Gauvin-Ndiay {\it et al.}\cite{Gauvin-Ndiaye_PRB2018} for double perovskites does not seem valid for the double-double perovskite CaMnCrSbO$_{6}$, a relatively more complex system. 
Hence, we need to regard each Mn$^{2+}$ spin in a unit cell as belonging to a different sublattice.
Thus, there are four Mn$^{2+}$ sublattices and one Cr$^{3+}$ sublattice in the unit cell.
We consider the entire spin lattice to be consisting of $2N$ spins, with $N$ Cr$^{3+}$ spins and $N$/4 spins in each magnetic sublattice of Mn$^{2+}$ spins. We denote the $z$-components of the four different Mn$^{2+}$ sublattice spins and the Cr$^{3+}$ spins as $S_{M1}^{z}$, $S_{M2}^{z}$, $S_{M3}^{z}$, $S_{M4}^{z}$ and $S_{C}^{z}$, respectively. It may be noted that $S_{M1}^{z}$ and $S_{M4}^{z}$ correspond to Mn(1) atoms, while $S_{M2}^{z}$ and $S_{M3}^{z}$ correspond to Mn(2) atoms. Using the same notation of $J$`s as in Table 4 and Fig.~\ref{exchange_lattice}, the exchange Hamiltonian of the equation (1) is now written as,
\begin{equation}
\begin{split}
H  & = -J_{1}{\sum\limits_{i,j=1}^{N/4}{{{S_{M1i}^{z}}{S_{M2j}^{z}}}}}-J_{1}{\sum\limits_{i,j=1}^{N/4}{{{S_{M3i}^{z}}{S_{M4j}^{z}}}}}-
J_{1a}'{\sum\limits_{i,j=1}^{N/4}{{{S_{Ci}^{z}}{(S_{M1j}^{z}+S_{M4j}^{z})}}}}-J_{1b}'{\sum\limits_{i,j=1}^{N/4}{{{S_{Ci}^{z}}{(S_{M2j}^{z}+S_{M3j}^{z})}}}} \\
& - J_{2}{\sum\limits_{i,j=1}^{N/4}{{{S_{M1i}^{z}}{S_{M3j}^{z}}}}}-J_{2}{\sum\limits_{i,j=1}^{N/4}{{{S_{M2i}^{z}}{S_{M4j}^{z}}}}}-J_{2}'{\sum\limits_{i,j=1}^{N}{{{S_{Ci}^{z}}{S_{Cj}^{z}}}}}
\end{split}
\end{equation}
In the mean field approximation, we substitute $S_{ai}^{z}$$=$($<$$S_{ai}^{z}$$>$+${\delta}$$S_{ai}^{z}$) ($a$$=$$M1$, .., $M4$ and $C$), where ${\delta}$$S_{ai}^{z}$=$S_{ai}^{z}$-$<$$S_{ai}^{z}$$>$ and neglect the second order terms (${\delta}$$S_{ai}^{z})^{2}$. The average sublattice magnetization is defined as, 
$M_{k}$$=$$(g{\mu}_{B}N/4)$$<$$S_{Mk}^{z}$$>$; $M_{C}$$=$$g{\mu}_{B}N<S_{C}^{z}>$
where $M_{k}$($k$ = 1 to 4) correponds to the sub-lattice magnetization of the four Mn$^{2+}$ spins, while $M_{C}$ is the sub-lattice magnetization of Cr$^{3+}$ spins.
 Using the mean field approximation (hereafter dropping the limits of summation), we obtain the mean field Hamiltonian,
\begin{equation}
\begin{split}
H_{MF} & = \frac{-J_{1}}{g{\mu}_{B}N}\left(2M_{2}{\sum S_{M1i}^{z}}+2M_{1}{\sum S_{M2i}^{z}}+2M_{4}{\sum S_{M3i}^{z}}+2M_{3}{\sum S_{M4i}^{z}}\right) \\
& \frac{-J_{2}}{g{\mu}_{B}N}\left(4M_{3}{\sum S_{M1i}^{z}}+4M_{4}{\sum S_{M2i}^{z}}+2M_{1}{\sum S_{M3i}^{z}}+2M_{2}{\sum S_{M4i}^{z}}\right) \\
& -\frac{1}{g{\mu}_{B}N}\left(J_{1a}'(M_{1}+M_{4})+J_{1b}'(M_{2}+M_{3})\right){\sum S_{Ci}^{z}} \\
& -\frac{M_{C}}{g{\mu}_{B}N}\left(J_{1a}'({\sum S_{M1i}^{z}}+{\sum S_{M4i}^{z}})+J_{1b}'({\sum S_{M2i}^{z}}+{\sum S_{M3i}^{z}})\right) \\
& -\frac{24J_{2}'M_{C}}{g{\mu}_{B}N}{\sum S_{Ci}^{z}} \\
\end{split}
\end{equation}
In terms of effective mean fields ($h_{M}$ or $h_{C}$) acting on each sub-lattice atom,
\begin{equation}
\label{mean_field_hamiltonian}
H_{MF}  = 
{{{h_{M1}}\sum{S_{M1i}^{z}}}}+{{{h_{M2}}\sum{S_{2i}^{z}}}}+{{{h_{M3}}\sum{S_{3i}^{z}}}}+{{{h_{M4}}{\sum S_{M4i}^{z}}}}+{{{h_{C}}{\sum{S_{Ci}^{z}}}}}
\end{equation}
where the exchange(mean) field acting on each sublattice is,
 \begin{equation}
\begin{split}
h_{M1} & = \frac{-1}{g{\mu}_{B}N'}(2J_{1}M_{2}+4J_{2}M_{3}+J_{1a}'M_{C}) \\ 
h_{M2} & = \frac{-1}{g{\mu}_{B}N'}(2J_{1}M_{1}+4J_{2}M_{4}+J_{1b}'M_{C}) \\ 
h_{M3} & = \frac{-1}{g{\mu}_{B}N'}(2J_{4}M_{2}+4J_{1}M_{3}+J_{1b}'M_{C}) \\ 
h_{M4} & = \frac{-1}{g{\mu}_{B}N'}(2J_{1}M_{3}+4J_{2}M_{2}+J_{1a}'M_{C}) \\ 
h_{C} & = \frac{-1}{g{\mu}_{B}N}(J_{1a}'(M_{1}+M_{4})+J_{1b}'(M_{2}+M_{3})+24J_{2}'M_{C}) \\ 
\end{split}
\end {equation} 
where $N'$$=$$N$/4. The total Hamiltonian, which comprises of the Zeeman and mean field terms is,
\begin{equation}
\begin{split}
H_{total} & = -g{\mu}_{B}\{H_{0}+\frac{h_{M1}}{g{\mu}_{B}}\}\sum S_{M1i}^{z}-g{\mu}_{B}\{H_{0}+\frac{h_{M2}}{g{\mu}_{B}}\}\sum S_{M2i}^{z}\\
& -g{\mu}_{B}\{H_{0}+\frac{h_{M3}}{g{\mu}_{B}}\}\sum S_{M3i}^{z}-g{\mu}_{B}\{H_{0}+\frac{h_{M4}}{g{\mu}_{B}}\}\sum S_{M4i}^{z} \\
& -g{\mu}_{B}\{H_{0}+\frac{h_{C}}{g{\mu}_{B}}\}\sum S_{Ci}^{z}
\end{split}
\end{equation}
where $H_{0}$ is the applied magnetic field.
The term within the curly brackets denoted as $h_{a}^{eff}$ ($a$=M1,..M4 and Cr), corresponds to the total effective field. 
In terms of the effective fields ahead in equation (19), the thermal average of sub-lattice magnetization for the Mn$^{2+}$ and Cr$^{3+}$ spins is,
\begin{equation}
\begin{split}
M_{Mk} & = gS_{M}{\mu}_{B}N'B_{s}\left(\frac{gS_{M}{\mu}_{B}h_{Mk}^{eff}}{k_{B}T}\right)(k=1,..,4) \\
M_{C} & = gS_{C}{\mu}_{B}NB_{s}\left(\frac{gS_{C}{\mu}_{B}h_{C}^{eff}}{k_{B}T}\right)
\end{split}
\end{equation}
where $B_{s}(x)$(=$\frac{2S+1}{2S}$coth($\frac{2S+1}{2S}$$x$)-$\frac{1}{2S}$coth($\frac{x}{2S}$)) is the Brillouin function.
For $x$${<}$${<}$1, i.e. when ${\mu}_{B}h_{a}^{eff}$${<}$${<}k_{B}T$ ($a$: M or C) which is true for large temperatures, the sub-lattice magnetization for Mn$^{2+}$ and Cr$^{3+}$ spins can be written as,  
\begin{equation}
\begin{split}
M_{k} & = g^{2}{\mu}_{B}^{2}N'S_{M}(S_{M}+1)\frac{h_{Mk}^{eff}}{3k_{B}T} \\
M_{C} & = g^{2}{\mu}_{B}^{2}NS_{C}(S_{C}+1)\frac{h_{C}^{eff}}{3k_{B}T} \\
\end{split}
\end{equation}
Setting $H_{0}$$=$0, from equation (18)-(20), we obtain a system of linear equations,
\begin{equation}
\begin{split}
M_{1} & = \frac{S_{M}(S_{M}+1)}{3k_{B}T}(2J_{1}M_{2}+4J_{2}M_{3}+J_{1a}'M_{C}) \\
M_{2} & = \frac{S_{M}(S_{M}+1)}{3k_{B}T}(2J_{1}M_{1}+4J_{2}M_{4}+J_{1b}'M_{C}) \\
M_{3} & = \frac{S_{M}(S_{M}+1)}{3k_{B}T}(2J_{4}M_{2}+4J_{1}M_{3}+J_{1b}'M_{C}) \\
M_{4} & = \frac{S_{M}(S_{M}+1)}{3k_{B}T}(2J_{1}M_{3}+4J_{2}M_{2}+J_{1a}'M_{C}) \\
M_{C} & = \frac{S_{C}(S_{C}+1)}{3k_{B}T}(J_{1a}'(M_{1}+M_{4})+J_{1b}'(M_{2}+M_{3})+24J_{2}'M_{C}) \\
\end{split}
\end{equation}
The transition temperature $T_{C}$ is obtained from the homogeneous solution of equation (21) i.e.
 \begin{equation}
\label{determinant}
\begin{vmatrix}
 1 & \frac{-2J_{1} \bar S_{1}}{y} & \frac{-4J_{2} \bar S_{1}}{y} & 0 & \frac{-J_{1a}' \bar S_{1}}{y} \\
& \\
\frac{-2J_{1} \bar S_{1}}{y}   & 1 & 0  &  \frac{-4J_{2} \bar S_{1}}{y} & \frac{-J_{1b}' \bar S_{1}}{y} \\
&\\
\frac{-4J_{2} \bar S_{1}}{y} & 0 & 1 & \frac{-2J_{1} \bar S_{1}}{y} & \frac{-J_{1b}' \bar S_{1}}{y} \\
& \\
0 & \frac{-4J_{2} \bar S_{1}}{y} & \frac{-2J_{1} \bar S_{1}}{y} & 1 & \frac{-J_{1a}' \bar S_{1}}{y} \\
& \\
\frac{-J_{1a}' \bar S_{2}}{y} & \frac{-J_{1b}' \bar S_{2}}{y} & \frac{-J_{1b}' \bar S_{2}}{y} & \frac{-J_{1a}' \bar S_{1}}{y}  & 1-\frac{24J_{2}' \bar S_{2}}{y} \\
\end{vmatrix}
= 0
\end{equation}
where, $y$$=$3$k_{B}$$T_{C}$, ${\bar S_{1}}$$=$$S_{M}(S_{M}+1)$, ${\bar S_{2}}$$=$$S_{C}(S_{C}+1)$.
Solving equation (22), we obtain $T_{C}$$=$ 99\,K, which is closer to the experimental value of 49\,K, while the two sublattice model yields a considerably higher value of ${\sim}$ 188 \,K.
\section{Summary and conclusions}
We have investigated thoroughly the electronic and magnetic structure along with magnetic exchange interactions in the recently reported double-double perovskite CaMnCrSbO$_{6}$.
Amongst the various arrangements of Mn$^{2+}$ and Cr$^{3+}$ spins, the lowest energy corresponds to ferrimagnetic configuration, in agreement with the experimental observation.
For $U_{eff}$$=$0 eV, i.e. without inclusion of electronic correlations, a band gap of ${\sim}$0.7 eV is obtained which increases on increasing the $U_{eff}$, thereby suggesting that the system has a Mott-Hubbard like character.
Our non-collinear calculations reveal that the combined Mn$^{2+}$/Cr$^{3+}$ spins show anisotropy in the crystallographic $x$-direction, thus resolving the ambiguity (in the experimental results) about the preferred direction of the spins.
The inclusion of the anti-site disorder explains the reduced value of the Mn$^{2+}$ magnetic moments in this compound. 
\newline
The various competing exchange interactions in the system have been evaluated by mapping the total energies obtained from DFT to the Ising model.
%
%
%
The nearest neighbour (along $c$-direction) Mn(1)-O-Mn(2) superexchange interactions are anti-ferromagnetic and weaker than the ferromagnetic Cr-O-O-Cr super-superexchange interactions, while the in-plane Mn(1)-O-Mn(2) superexchange is the weakest in the lattice.
The Mn-O-Cr interactions are also predominantly antiferromagnetic.
A simple two sublattice model (Mn(1)$^{2+}$ and Mn(2)$^{2+}$ spin-up, Cr$^{3+}$ spin-down) thus cannot explain the magnetism of this system.
Due to difference in the ligand symmetry between Mn(1) and Mn(2), the Mn(1)-O-Cr superexchange interaction is stronger than the Mn(2)-O-Cr interaction, thereby establishing a ``ferrimagnetic" order.
\newline
The origins and the strengths of the various superexchange interactions have been calculated in terms of the hopping between the Wannier orbitals of Mn(1), Mn(2) and Cr via the intermediate O $p$-orbitals. Using a 3-site Hubbard model with singly occupied $d$-orbital at both TM sites and a doubly occupied O $p$-orbital, we obtain expression for $J_{AFM}$. 
%
%
 The formula predicts correct relative strengths for the various anti-ferromagnetic interactions. While the maximum $J_{AFM}$ for Mn(1)-O-Mn(2) (out of plane and in-plane) is closer to the DFT results, the maximum $J_{AFM}$ for Mn-O-Cr superexchange is overestimated by an order of magnitude.
Similarly, considering hopping via two different $p$-orbitals, we obtain expression for $J_{FM}$, which also exists in a 90$^{o}$ superexchange. Though much smaller than $J_{AFM}$, it can result in an overall decrease in the antiferromagnetic superexchange strengths.
%
Using a 4-site Hubbard model, we obtain an expression for the Cr-O-O-Cr super-superexchange ${\tilde J}_{SS}$, which is found to be ferromagnetic, with a maximum value in close agreement with the DFT results. 
We show that considering a set of five magnetic sub-lattices, using the mean field theory one gets a Curie temperature of 99\,K, which is closer to the experimental value of 49\,K unlike the two-sublattice model which yields a higher Curie temperature.

\section{Acknowledgments}
The authors wish to thank the Institute Compute Centre of IIT Roorkee for providing the computational facility. 

\section{Appendix} 
{\bf Appendix A.} AFM Super-exchange interaction
\newline
 The Hubbard Hamiltonian for the first order neighbour superexchange between two TM ions and a single oxygen ion can be written as,
\setcounter{equation}{0}
\renewcommand{\theequation}{A\arabic{equation}}
 \begin{equation}
H = -\sum_{{\sigma}={\uparrow},{\downarrow}}[{\epsilon}_{d1}{\hat n_{1{\sigma}}}+{\epsilon}_{d2}{\hat n_{2{\sigma}}}+{\epsilon}_{p}{\hat n_{p{\sigma}}}-\sum_{i=1,2}t_{pdi}(d_{i{\sigma}}^{+}c_{p{\sigma}}+c_{p{\sigma}}^{+}d_{i{\sigma}})]+U_{1}{\hat n_{1{\uparrow}}}{\hat n_{1{\downarrow}}}+U_{2}{\hat n_{2{\uparrow}}}{\hat n_{2{\downarrow}}}
\end{equation}
where ${\epsilon}_{da}$($a$=1,2) correspond to the on-site energies of the $d$-orbitals of the two TM ions under consideration, ${\epsilon}_{p}$ is the on-site energy of the O $p$-orbital, $t_{pd}$ is the hopping integral between the TM $d$- O $p$-orbitals, $U_{1}$ and $U_{2}$ are the Hubbard $U$ parameters for the first and second TM ions respectively.
Due to the distinct TM ions, (Mn(1)$^{2+}$, Mn(2)$^{2+}$ and Cr$^{3+}$), the various parameters (on-site, hopping and Hubbard) are different, hence we use additional index $i$.
The ground state comprises of singly occupied $d$-orbitals (parallel or anti-parallel). The Hamiltonian matrix for the ferromagnetic (spin-triplet) case is given as,
\begin{equation}
H_{FM}= 
\begin{bmatrix}
 0 & t_{pd1} &  t_{pd2}  \\
t_{pd1}& {\tilde U}_{1} & 0  \\
t_{pd2}& 0 & {\tilde U}_{2} \\
\end{bmatrix}
\end{equation}
The first term which corresponds to ${\epsilon}_{d1}$+${\epsilon}_{d2}$+2${\epsilon}_{p}$ is set to be 0. We define, ${\Delta}_{pdi}$$=$${\epsilon}_{di}$-${\epsilon}_{p}$ as the charge transfer energy. In the above matrix, we have made the following abbreviations,
${\tilde U}_{1}$$=$$U_{1}$+${\Delta}_{pd1}$, ${\tilde U}_{2}$$=$$U_{2}$+${\Delta}_{pd2}$.

The corresponding matrix for anti-ferromagnetic (spin singlet) case is,
\begin{equation}
H_{AFM}= 
\begin{bmatrix}
 0 & 0 &t_{pd}^{1} & t_{pd}^{2} & 0 & 0 & 0 & 0 & 0\\
0 & 0 & 0 & 0 & t_{pd}^{1} & t_{pd}^{2} & 0 & 0 & 0 \\
t_{pd}^{1} &0 & {\tilde U}_{1} & 0& 0 & 0 & -t_{pd}^{2} & 0 & -t_{pd}^{2}\\
t_{pd}^{2} &0 & 0 & {\tilde U}_{2} & 0  &0 & 0 & -t_{pd}^{1} & -t_{pd}^{1} \\
0 & t_{pd}^{1} & 0 & 0 & {\tilde U}_{1}  & 0 & t_{pd}^{2} & 0 & t_{pd}^{2}  \\
0 & t_{pd}^{1} & 0  & 0  & 0 & {\tilde U}_{2} & 0 & t_{pd}^{1} & t_{pd}^{1}\\ 
0 & 0 & -t_{pd}^{2} & 0 & t_{pd}^{2} & 0 & U_{1}+{\Delta}{\epsilon} & 0 & 0  \\
0 & 0 & 0 & -t_{pd}^{1} & 0 & t_{pd}^{1} & 0 & U_{2}-{\Delta}{\epsilon} & 0 \\
0 & 0 & -t_{pd}^{2} & -t_{pd}^{1} & t_{pd}^{2} & t_{pd}^{1} & 0  & 0 & {\bar U} \\
\end{bmatrix}
\end{equation}
 where, ${\bar U}$$=$${\tilde U}_{1}$+${\tilde U}_{2}$ and ${\Delta}{\epsilon}$$=$${\epsilon}_{d1}$-${\epsilon}_{d2}$.
Using the method of downfolding, we obtain an effective Hamiltonian,
\begin{equation}
H_{eff} = -T_{01}H_{11}^{-1}T_{10}-T_{01}H_{11}^{-1}T_{12}H_{22}^{-1}T_{21}H_{11}^{-1}T_{10}
\end{equation}
For the FM case, only the first term in equation (A4) exists. 
 \begin{equation}
H_{eff}^{FM} = -\frac{t_{pd1}^{2}}{{\tilde U}_{1}}-\frac{t_{pd2}^{2}}{{\tilde U}_{2}}
 \end{equation}
 \begin{equation}
\begin{split}
H_{eff}^{AFM} & = \left(-\frac{t_{pd}^{1}}{{\tilde U}_{1}}-\frac{t_{pd}^{2}}{{\tilde U}_{2}}\right) \begin{bmatrix}
1 & 0\\
0 & 1\\
\end{bmatrix}
-(t_{pd1}t_{pd2})^{2}\left[\frac{1}{{\tilde U}_{1}^2(U_{1}+{\Delta}{\epsilon})}+\frac{1}{{\tilde U}_{2}^2(U_{2}-{\Delta}{\epsilon})}+ \frac{1}{{\bar U}}\left(\frac{1}{{\tilde U}_{1}}+\frac{1}{{\tilde U}_{2}}\right)^{2}\right] \\
{\times} & \begin{bmatrix} 
 1 & -1\\
 -1 & 1\\
\end{bmatrix}
\end{split}
\end{equation}
Subtracting (A5) from (A6) yields the antiferromagnetic exchange constant 
\begin{equation}
J_{AFM} = -2(t_{pd1}t_{pd2})^{2}\left[\frac{1}{{\tilde U}_{1}^2(U_{1}+{\Delta}{\epsilon})}+\frac{1}{{\tilde U}_{2}^2(U_{2}-{\Delta}{\epsilon})}+ \frac{1}{{\bar U}}\left(\frac{1}{{\tilde U}_{1}}+\frac{1}{{\tilde U}_{2}}\right)^{2}\right]
\end{equation}
The sign of antiferromagnetic exchange is negative.
\newline
{\bf Appendix B.} FM superexchange interaction
\newline
The ferromagnetic superexchange which involves two $p$-orbitals of the intermediate O$^{2-}$ ion can be described by the Hamiltonian,
\setcounter{equation}{0}
\renewcommand{\theequation}{B\arabic{equation}}
 \begin{equation}
\begin{split}
H & = \sum_{{\sigma}={\uparrow},{\downarrow}}[{\epsilon}_{d1}{\hat n_{1{\sigma}}}+{\epsilon}_{d2}{\hat n_{2{\sigma}}}+{\epsilon}_{p}({\hat n_{p1{\sigma}}}+{\hat n_{p2{\sigma}}})-J_{p}{\hat n_{p1,{\sigma}}}{\hat n_{p2,-{\sigma}}} \\
&-\sum_{i=1,2}t_{pdi}(d_{i{\sigma}}^{+}c_{p{\sigma}}+c_{p{\sigma}}^{+}d_{i{\sigma}})] 
 + U_{1}{\hat n_{1{\uparrow}}}{\hat n_{1{\downarrow}}}+U_{2}{\hat n_{2{\uparrow}}}{\hat n_{2{\downarrow}}} \\
\end{split}
\end{equation}
The Hamiltonian now involves occupancy of the two $p$-orbitals, while $J_{p1,p2}$ corresponds to the exchange term between the two $p$-orbitals. The on-site repsulsion at the O$^{2-}$ ion is neglected. Similarly, no hopping occurs between the two $p$-orbitals of the same anion.
The matrix representation for the triplet ground state is,
\begin{equation}
H_{FM}= \begin{bmatrix}
 0 & t_{pd1} &  t_{pd2} & 0 \\
t_{pd1}& {\tilde U}_{1} & 0 & t_{pd1}  \\
t_{pd2}& 0 & {\tilde U}_{2} & t_{pd2} \\
0 & t_{pd1} & t_{pd2} & {\tilde U}-J_{p} \\ 
\end{bmatrix}
\end{equation}

The corresponding Hamiltonian matrix for the singlet state is 
\begin{equation}
H_{AFM}= \begin{bmatrix}
 0 & 0 &t_{pd}^{1} & 0 & t_{pd}^{2} & 0 & 0 & 0 \\
0 & 0 & 0 & t_{pd}^{1} & 0 & t_{pd}^{2} & 0 & 0 \\
t_{pd}^{1} &0 & {\tilde U}_{1} & 0& 0 & 0 & -t_{pd}^{2} & 0 \\
0 & t_{pd}^{2} & 0 & {\tilde U}_{2} & 0  &0 & 0 & -t_{pd}^{1}  \\
t_{pd}^{1} & 0 & 0 & 0 & {\tilde U}_{1}  & 0 & t_{pd}^{2} & 0  \\
0 & t_{pd}^{1} & 0  & 0  & 0 & {\tilde U}_{2} & 0 & t_{pd}^{1} \\ 
0 & 0 & -t_{pd}^{2} & 0 & -t_{pd}^{2} & 0 &  {\tilde U}  & -J_{p}  \\
0 & 0 & 0 & -t_{pd}^{1} & 0 & t_{pd}^{1} & -J_{p} & {\tilde U}  \\
\end{bmatrix}
\end{equation}
The effective Hamiltonian for the triplet is given by,
\begin{equation}
H_{eff}^{FM} = -\left(\frac{t_{pd1}^{2}}{{\tilde U}_{1}}+\frac{t_{pd2}^{2}}{{\tilde U}_{2}}\right)-\frac{\left(\frac{t_{pd1}^{2}}{{\tilde U}_{1}}+\frac{t_{pd2}^{2}}{{\tilde U}_{2}}\right)^{2}}{{\tilde U}-J_{p}}
 \end{equation}
while for the singlet, 
\begin{equation}
H_{eff}^{AFM}=-\left(\frac{t_{pd1}^{2}}{{\tilde U}_{1}}+\frac{t_{pd2}^{2}}{{\tilde U}_{2}}\right)-\frac{\left({t_{pd1}^{2}}+{t_{pd2}^{2}}\right)^{2}}{J_{p}^{2}-{\tilde U}^{2}} 
\begin{bmatrix}
\frac{{\tilde U}}{{\tilde U}_{1}^2} & \frac{-J_{p}}{{\tilde U}_{1}{\tilde U}_{2}} \\
\frac{-J_{p}}{{\tilde U}_{1}{\tilde U}_{2}} & \frac{{\tilde U}}{{\tilde U}_{2}^{2}} \\
\end{bmatrix}
\end{equation}
The exchange interaction strength for the ferromagnetic superexchange is obtained as,
\begin{equation}
J_{FM} = \frac{\left(\frac{t_{pd1}^{2}}{{\tilde U}_{1}}+\frac{t_{pd2}^{2}}{{\tilde U}_{2}}\right)^{2}}{{\tilde U}-J_{p}}-\frac{\left({t_{pd1}^{2}}+{t_{pd2}^{2}}\right)^{2}}{J_{p}^{2}-{\tilde U}^{2}} 
\begin{bmatrix}
\frac{{\tilde U}}{{\tilde U}_{1}^2} & \frac{-J_{p}}{{\tilde U}_{1}{\tilde U}_{2}} \\
\frac{-J_{p}}{{\tilde U}_{1}{\tilde U}_{2}} & \frac{{\tilde U}}{{\tilde U}_{2}^{2}} \\
\end{bmatrix}
\end{equation}
The sign of $J_{FM}$ is positive.
\newline
{\bf Appendix C.} Super-superexchange interaction
\newline
The exchange interaction between the Cr$^{3+}$ spins occurs via hopping through two O$^{2-}$ ions, thus the Hamiltonian can be written as,
\setcounter{equation}{0}
\renewcommand{\theequation}{C\arabic{equation}}
\begin{equation}
H = -\sum_{{\sigma}={\uparrow},{\downarrow}}\sum_{i=1,2}[{\epsilon}_{di}{\hat n_{i{\sigma}}}+{\epsilon}_{pi}{\hat n_{pi{\sigma}}}+U_{i}{\hat n_{i{\uparrow}}}{\hat n_{i{\downarrow}}}]-t_{pd}(d_{1{\sigma}}^{+}c_{p1{\sigma}}+d_{2{\sigma}}^{+}c_{p2{\sigma}})-t_{pp}c_{p2{\sigma}}^{+}c_{p1{\sigma}}+\text{h.c}
\end{equation}
In the above equation, $t_{pp}$ corresponds to the hopping amplitude between the two O$^{2-}$ $p$-orbitals.
The matrix corresponding to the triplet state is, 
\begin{equation}
H_{FM}= 
\begin{bmatrix}
 0 & t_{pd} &t_{pd} & 0 & 0 & 0  \\
t_{pd} & {\tilde U} & t_{pd} & t_{pp} & 0 & t_{pd}  \\
t_{pd} &t_{pd} & {\tilde U} & 0& t_{pp} & t_{pd}  \\
0 &t_{pp} & 0 & {\tilde U} & 0  &0  \\
0 & 0 & t_{pp} & 0 & {\tilde U} & 0     \\
0 & t_{pd} & t_{pd} & 0  & 0 & 2{\tilde U}  \\ 
\end{bmatrix}
\end{equation}
After downfolding (using equation (A4)), the effective Hamiltonian for the triplet state is,
\begin{equation}
H_{eff}^{FM} = -\frac{2t_{pd}^2}{{\tilde U}+t_{pd}}-\frac{2t_{pd}^2(t_{pd}^2+t_{pp}^2)}{{\tilde U}({\tilde U}+t_{pd})^2}\begin{bmatrix}
1 & 1\\
1 & 1\\
\end{bmatrix}
\end{equation} 
The results for the singlet state are more complex yielding a 16x16 matrix, which in block form can be written as,
\begin{equation}
H_{AFM}= 
\begin{bmatrix}
 H_{00} & T_{01} &  T_{02} & T_{03} \\
T_{10}& H_{11} & T_{12} & T_{13}  \\
T_{20}& T_{21} & H_{22} & T_{23}  \\
T_{30}& T_{31} & T_{32} & H_{33}  \\
\end{bmatrix}
\end{equation}
$H_{00}$ is a 2x2 diagonal matrix corresponding to the ground state on-site energy 2${\epsilon}_{d}$+4${\epsilon}_{p}$ which is set to zero.
The other block matrices along the diagonal are,
\begin{equation}
H_{11}= 
\begin{bmatrix}
{\tilde U} & 0 & 0 & 0 & 0 & 0 & t_{pp} & 0 \\
0 & {\tilde U} & 0 & 0 & 0 & 0 & 0 & t_{pp}  \\
0 &0 & {\tilde U} & 0& t_{pp} & 0 & 0 & 0 \\
0 &0 & 0 & {\tilde U} & 0  &t_{pp} & 0 & 0  \\
0 & 0 & t_{pp} & 0 & {\tilde U}  & 0 & 0 & 0  \\
0 & 0 & 0  & t_{pp} & 0 & {\tilde U} & 0 & 0 \\ 
t_{pp} & 0 & 0 & 0 & 0 & 0 & {\tilde U} & 0  \\
0 & t_{pp} & 0 & 0 & 0 & 0 & 0 & {\tilde U} \\
\end{bmatrix},
\quad
H_{22}= 
\begin{bmatrix}
 U & 0\\
0 & U  \\
\end{bmatrix},
 \quad
H_{33}= 
\begin{bmatrix}
 2{\tilde U} & 0 & t_{pp} & t_{pp} \\
0 & 2{\tilde U} & -t_{pp} & -t_{pp} \\
t_{pp} & -t_{pp} & 2{\tilde U} & 0 \\
t_{pp} & -t_{pp} & 0 & 2{\tilde U} \\
\end{bmatrix}
\end{equation}
where ${\tilde U}$$=$$U$+${\Delta}_{pd}$. The off-diagonal block matrices are given below
 \begin{equation}
T_{10}= 
\begin{bmatrix}
t_{pd} & 0  \\
0 & t_{pd} \\
t_{pd} &0  \\
0 &t_{pd}   \\
0 & 0 \\
0 & 0 \\ 
0 & 0  \\
0 & 0 \\
\end{bmatrix},
\quad
T_{12}= 
\begin{bmatrix}
0 & 0  \\
0 & 0 \\
0 &0  \\
0 &0   \\
0 & t_{pd} \\
0 & -t_{pd} \\ 
t_{pd} & 0  \\
-t_{pd} & 0 \\
\end{bmatrix},
\quad
T_{13}= 
\begin{bmatrix}
 t_{pd} & 0 & 0 & 0 \\
0 & t_{pd} & 0 & 0 \\
 t_{pd} & 0 & 0 & 0 \\
0 & t_{pd} & 0 & 0 \\
0 & 0 & t_{pd} & 0 \\
0 & 0 & -t_{pd} & 0 \\
0 & 0 & 0 & t_{pd} \\
0 & 0 & 0 & -t_{pd} \\
\end{bmatrix}
\end{equation}
$T_{01}$$=$$T_{10}^{\dagger}$, $T_{21}$$=$$T_{12}^{\dagger}$ and $T_{31}$$=$$T_{13}^{\dagger}$. The rest of the block matrices are null matrices.
Using the method of downfolding, the effective Hamiltonian is given as,
\begin{equation}
H_{eff}^{AFM} = -T_{01}H_{11}^{-1}T_{10}-T_{01}H_{11}^{-1}T_{12}H_{22}^{-1}T_{21}H_{11}^{-1}T_{10}-T_{01}H_{11}^{-1}T_{13}H_{33}^{-1}T_{31}H_{11}^{-1}T_{10}
\end{equation}
The corresponding effective Hamiltonian of the singlet state is,
 \begin{equation}
H_{eff}^{AFM} = \frac{2t_{pd}^2{\tilde U}}{t_{pp}^2-{\tilde U}^{2}}-\frac{2t_{pd}^{4}t_{pp}^{2}}{U(t_{pp}^{2}-{\tilde U}^{2})^{2}}\begin{bmatrix}
1 & 1 \\
1 & 1\\ 
\end{bmatrix} + \frac{2t_{pd}^{4}{\tilde U}}{(t_{pp}-{\tilde U})^{3}(t_{pp}+{\tilde U})^{3}}\begin{bmatrix}
{\tilde U}^{2}+t_{pp}^{2} & t_{pp}^{2} \\
t_{pp}^2 & {\tilde U}^{2}+t_{pp}^{2} \\ 
\end{bmatrix}
\end{equation}
The final expression for the super-superexchange interaction strength, which is the difference between equation(A21) and (A16), is
\begin{equation}
J_{SS}=\frac{2t_{pd}^2(t_{pd}^2+t_{pp}^2)}{{\tilde U}({\tilde U}+t_{pd})^{2}}-\frac{2t_{pd}^{4}t_{pp}^{2}}{U{\tilde U}^{4}}-\frac{2t_{pd}^{4}}{{\tilde U}^{3}}+\frac{4t_{pd}^{4}t_{pp}^{2}}{{\tilde U}^{5}}
\end{equation}
The first terms in the R.H.S of equations (C3) and (C8) correspond to a direct exchange process and hence not considered. Equation (C9) is obtained after making the approximation $t_{pp}$ ${<}{<}$ ${\tilde U}$ in equation (C8).  
\section{References}
 \bibliography{CaMnCrSbO6}
\end{document}